\documentclass[graybox]{svmult}

\usepackage{mathptmx}       
\usepackage{helvet}         
\usepackage{courier}        
\usepackage{type1cm}        
\usepackage{makeidx}         
\usepackage{graphicx}        
\usepackage{multicol}        
\usepackage[bottom]{footmisc}

\makeindex             

\begin{document}

\title*{Laboratory experiments and numerical simulations on magnetic instabilities}
\author{F. Stefani, M. Gellert, Ch. Kasprzyk, 
A. Paredes, 
G. R\"udiger and M. Seilmayer}
\institute{F. Stefani, Ch. Kasprzyk, M. Seilmayer 
\at Helmholtz-Zentrum Dresden-Rossendorf, 
Bautzner Landstr. 400, 01328 Dresden, Germany, 
\email{F.Stefani@hzdr.de}
\and Marcus Gellert, A. Paredes, G. R\"udiger
 \at Leibniz-Institut f\"ur Astrophysik Potsdam, 
 An der Sternwarte 16, D-14482 Potsdam, Germany, \email{mgellert@aip.de} 
 }
%

%
\maketitle

\abstract{Magnetic fields of planets, stars and galaxies are generated 
by self-excitation in moving electrically conducting fluids. Once produced, 
magnetic fields can play an active role in cosmic structure formation 
by destabilizing rotational flows that would be otherwise hydrodynamically 
stable. For a long time, both hydromagnetic dynamo action as well as 
magnetically triggered flow instabilities had been the subject of purely 
theoretical research. Meanwhile, however, the dynamo effect has been
observed in large-scale liquid sodium experiments in Riga, 
Karls\-ruhe and Cadarache. In this paper, we summarize the 
results of some smaller
liquid metal experiments devoted to 
various magnetic instabilities such as the 
helical and the azimuthal magnetorotational instability, the 
Tayler instability, and
the different instabilities that appear in a
magnetized spherical Couette flow. 
We conclude with an outlook on a large scale 
Tayler-Couette experiment using liquid sodium, and
on the prospects to observe magnetically triggered 
instabilities of flows with positive shear.
}

\section{Introduction}
\label{sec:1}
Magnetic fields of planets and stars are known to be produced by the 
homogeneous dynamo effect \cite{Jones_2011,Wicht_Tilgner_2010}. After 
decades of mainly theoretical and numerical work, the last years have 
seen tremendous progress in complementary experimental studies devoted 
to a better understanding 
of self-excitation in homogeneous fluids 
\cite{Gailitis_2002,Lathrop_Forest_2011,Stefani_2008}. 
Following the pioneering Riga and Karlsruhe dynamo experiments
\cite{Gailitis_2000,Stieglitz_2001}, it was in particular 
the rich dynamics observed in the French von K\'arm\'an Sodium (VKS) 
experiment \cite{Berhanu_2010} that has provoked interest 
throughout the dynamo community.
The observed reversals, excursions, bursts, 
hemispherical fields etc. have boosted considerable research directed 
to a deeper physical understanding of the corresponding geomagnetic
phenomena 
\cite{Benzi_2010,Mori_2013,Petrelis_2009,Sorriso_2007,Stefani_2006a}.

While realistic ''bonsai'' models of planetary dynamos, with all 
dimensionless numbers matching those of planets, will never be 
possible in the laboratory \cite{Lathrop_Forest_2011}, 
there is still ongoing effort to explore the technical limits of dynamo 
experiments. 
This applies to the 3 m diameter spherical Couette experiment 
presently being spun at the University of Maryland 
\cite{Adams_2015,Zimmermann_2014}, 
to the 2 m diameter precession 
dynamo experiment \cite{Stefani_2012,Stefani_2015} 
that is under construction at 
Helmholtz-Zentrum Dresden-Rossendorf (HZDR), as well as to 
the 3 m diameter 
plasma dynamo experiment in Madison
\cite{Cooper_2014}. One of the 
characteristics of those ''second generation'' 
dynamo experiments is a larger degree of uncertainty 
of success. Whereas the Riga and Karlsruhe experiments 
had turned out to be well predictable by kinematic dynamo codes and 
simplified saturation models \cite{Stefani_2009} 
(and even the unexpected VKS dynamo results can be 
explained when correctly including the high permeability disks 
\cite{Giesecke_2010,Giesecke_2012, Nore_2016}), the outcomes of the 
Maryland, HZDR and Madison experiments are much harder to predict. 
In either case, this uncertainty 
follows directly from the ambition to construct a truly 
homogeneous dynamo, neither driven by pumps or propellers, 
nor influenced by guiding blades or gradients of magnetic 
permeability. This higher degree of freedom makes those 
experiments prone to emerging medium-size flow structures and waves. 
Actually, it is the influence of such waves that might foster, 
or inhibit, dynamo action in a much stronger way than any 
quasi-stationary analysis would suggest 
\cite{Reuter_2009,Tilgner_2008}. 
A closely related aspect here is the possibility of subcritical 
dynamo action under the influence of magnetic fields (as discussed, 
e.\,g., for rapidly rotating convection problems 
\cite{Dormy_2016,Sreenivasan_2011}).

Apart from this connection to dynamo action, 
instabilities and wave phenomena that appear 
under the common influence of rotation and magnetic fields are 
interesting in their own right, 
in particular
with respect to the 
angular momentum transport in planetary cores 
\cite{Petitdemange_2008,Petitdemange_2010}
and in active galactic nuclei and proto-planetary disks 
\cite{Balbus_2003} by 
virtue of the magneto-rotational instability (MRI). The fluctuations, 
arising from this and related magnetic flow instabilities, 
correlate and amplify the eddy viscosity of the fluid. The magnetic 
resistivity and the transport coefficients for temperature and 
mixing of chemicals, however, are less influenced by the
magnetic fluctuations. Consequently, 
the magnetic Prandtl number 
and the Schmidt number (i.\,e. the ratio of viscosity and 
diffusion coefficient) are enhanced under the presence of 
turbulent magnetic field components which are due to the instability 
of the magnetic background fields \cite{Paredes_2016}. This effect is supposed to 
explain the rigid rotation of the stellar interiors 
\cite{Spada_2016}, but may 
also play a role 
for the amplification of the fluid viscosity by magnetic 
instabilities in planetary cores.

It is for good reasons, therefore, that a number of medium-size 
liquid metal experiments are dedicated mainly to those 
instabilities, without ambition to reproduce the very 
dynamo effect. This applies, 
in particular, to the DTS experiment in Grenoble where a variety of 
magneto-inertial waves have been identified \cite{Schmitt_2013}, 
to the spherical Couette experiment in Maryland which has shown 
coherent velocity/magnetic field fluctuations 
quite reminiscent of MRI \cite{Sisan_2004}, as well as to the 
Taylor-Couette experiment in Princeton which has shown 
evidence for slow magneto-Coriolis waves \cite{Nornberg_2010} 
and a free-Shercliff layer instability \cite{Roach_2012}.
Despite these preliminary successes, the latter experiments have made 
it clear that for liquid metal flows the 
unambiguous identification 
of the standard MRI (SMRI), with a purely axial field being 
applied, 
is extremely complicated due to the key role of 
boundary effects (e.g. Ekman pumping) on the flow structure 
at those high Reynolds numbers ($\sim 10^6$) that are inevitably 
connected with the need to obtain magnetic Reynolds numbers 
of order unity.

These sobering prospects for studying MRI
in the lab suddenly brightened up with the numerical prediction 
\cite{Hollerbach_2005} of a very special, essentially 
inductionless, version of MRI whose onset does not depend 
on crossing certain critical magnetic Reynolds and Lundquist 
numbers, but only on crossing certain critical 
Reynolds and Hartmann numbers, 
which makes its experimental identification much easier. Since, in 
its axisymmetric form initially studied, this MRI version requires
the application of axial and azimuthal magnetic field of comparable 
strengths, it was coined helical MRI (HMRI) 
\cite{Liu_2006}. 
In 2006, a swiftly designed experiment at the PROMISE facility at HZDR 
had given first evidence for the onset of this HMRI, with roughly 
correct wave frequencies observed in the predicted parameter regions 
of the Hartmann number \cite{Stefani_2006,Stefani_2007}. In 2009,  
an improved version of this experiment - using split end-rings installed 
at the top and bottom of the cylinder in order to minimize the global 
Ekman pumping - allowed to observe the onset and cessation of HMRI for a 
number of parameter variations in much better agreement with numerical 
predictions \cite{Stefani_2009a}. After some preceding dispute about 
the (noise-triggered) convective or global character of the 
observed instability \cite{Liu_2007,Priede_2009}, the 
results gave now strong arguments in favour of the latter.

In parallel with these experimental works, Hollerbach et al. 
\cite{Hollerbach_2010} had identified a further induction-less 
MRI version that appears for strongly 
dominant azimuthal fields in form of a 
non-axisymmetric ($m=1$) mode, which is now called azimuthal MRI (AMRI). 
When further relaxing the demand that the azimuthal field should be 
current-free, one enters the realm of current-driven instabilities, 
including the Tayler instability (TI) \cite{Tayler_1973}, a kink-type 
instability in current-carrying conductors, whose ideal 
counterpart has been known for a long time from plasma z-pinch 
experiments \cite{Bergerson_2006}. Interestingly, the TI has 
been intensely 
discussed as a main ingredient of the 
Tayler-Spruit dynamo model 
\cite{Spruit_2002,Gellert_2008,Ruediger_2011}).
The first experimental investigation of the AMRI and the TI was the 
central goal of the project within DFG focus programme ''PlanetMag''. 
The corresponding numerical and 
experimental results for cylindrical geometry will
be the content of section Sect.~\ref{sec:2}.

A further topic was related to the different types of 
non-axisymmetric instabilities 
in spherical Couette flow that appear in dependence on the strength 
of an applied axial magnetic field \cite{Hollerbach_2009}.
Corresponding numerical predictions, and first experimental results,
will be reported in Sect.~\ref{sec:3}.

The paper closes with an outlook on the large-scale combined
MRI/TI experiment as it is planned in the framework of the DRESDYN
project at Helmholtz-Zentrum Dresden-Rossendorf,
and on the prospects to observe magnetically triggered 
instabilities of flows with positive shear.

\section{Instabilities in cylindrical geometry}
\label{sec:2}
In this section, we will present various numerical and 
experimental results on magnetically triggered flow 
instabilities in cylindrical geometry. These comprise
the AMRI and the TI, which are both non-axisymmetric instabilities
with azimuthal wavenumber $m=1$ that appear under the influence
of a purely (or dominantly) azimuthal magnetic field.
While AMRI draws its energy from the shear of the flow
and requires the current-free azimuthal field only 
as a trigger, the kink-type TI draws its energy from the electric 
current that is passing through the fluid. The axisymmetric HMRI, 
which becomes dominant when the azimuthal magnetic field is 
complemented by a vertical magnetic 
field of comparable magnitude, will also be discussed.

\subsection{Theory and Numerics}
\label{subsec:2-1}

During the last decade, the theory 
of HMRI, AMRI and TI has been developed in a rather 
comprehensive manner. 
This development was ignited by the observation of
Hollerbach and R\"udiger \cite{Hollerbach_2005}
that the combination of an axial and an azimuthal
magnetic field leads to an essentially 
inductionless  version of the axisymmetric 
MRI which does not scale with the 
magnetic Reynolds and the Lundquist 
number, but rather with the  Reynolds and 
Hartmann number. 
It is important to note that 
SMRI and HMRI are continuously and monotonically connected
\cite{Hollerbach_2005},
although the transition involves an exceptional point of the
spectrum where the slow magneto-Coriolis mode and the
inertial mode coalesce \cite{Kirillov_2010}.
Later it was also shown that the scaling properties of 
AMRI are essentially the same as those of HMRI
\cite{Kirillov_2012,Kirillov_2014}.

These 
instabilities can be treated with analytical and numerical 
methods of increasing complexity. In the following,
we will present in some detail the short-wavelength, or
Wentzel-Kramers-Brillouin (WKB) method. 
It allows  for analytical 
solutions in quite a couple of circumstances, and 
it provides easily a general overview about many 
parameter dependencies
and the transitions between different instabilities. 
Then we give  illustrating examples of 
1D modal stability analysis and of 3D simulations which
are necessary for quantitative predictions 
of real-world experiments. We also discuss some
recent results explaining turbulent angular momentum transport
and turbulent diffusion in stars in a consistent manner.

\subsubsection{Basic equations}
\label{subsubsec:2-1-1}

For the direct numerical simulations the MHD equations for an 
incompressible and electrically conduction fluid are solved. 
These are the coupled Navier-Stokes equation
for the velocity field $\bf{u}$ and the induction equation
for the magnetic field $\bf{B}$,
\begin{eqnarray}\label{eq_ns}
&\frac{\partial \bf{u}}{\partial t}+
{\bf{u}} \cdot \nabla {\bf{u}}-\frac{1}{\mu_0 \rho}{\bf{B}}\cdot 
\nabla {\bf{B}} +\frac{1}{\rho} \nabla P-
\nu \nabla^2 {\bf{u}}=0,&\\
&\frac{\partial {\bf{B}}}{\partial t}+{\bf{u}} \cdot \nabla 
{\bf{B}} - {\bf{B}} \cdot \nabla {\bf{u}}- \eta \nabla^2 {\bf{B}}=0,&
\end{eqnarray}
where $P=p+{\bf{B}}^2/(2\mu_0)$ is the total pressure, $\rho$
the density, $\nu$ the kinematic
viscosity, $\eta=(\mu_0 \sigma)^{-1}$ the magnetic diffusivity,
$\sigma$ the conductivity of the fluid,
and $\mu_0$ the magnetic permeability constant.
This set is complemented by the continuity equation for
incompressible flows and the solenoidal condition for the 
magnetic induction:
\begin{eqnarray}\label{eq_cs}
 \nabla \cdot {\bf{u}} & = & 0,\\
 \nabla \cdot {\bf{B}} & = & 0.
\end{eqnarray}

This equation system can also be written in dimensionless
form using the  Reynolds number 
${\rm Re}=\Omega_i L^2/\nu$, the Hartmann number
${\rm Ha} = B_0 L/\sqrt{\mu_0\rho\nu\eta}$, and 
the magnetic Prandtl number ${\rm Pm}=\nu/\eta$, 
where $B_0$ denotes the external field 
amplitude, $\Omega_i$ the angular frequency of an inner cylinder, 
and $L$ the gap width $L=r_o-r_i$ 
between an inner radius $r_i$ and an outer radius $r_o$
 
To study
the mixing properties the diffusion equation for 
a passive scalar
\begin{equation}\label{eqn:eq_tdiff}
\frac{\partial C}{\partial t} + \nabla \cdot 
\left({\bf{u}} C \right) =\frac{1}{\rm Sc} \Delta C
\end{equation}
is additionally solved. Here ${\rm Sc}=\nu/D^\ast$ is the Schmidt 
number and $D^\ast=D_{\rm mol}+D_{\rm turb}$ the effective 
diffusion coefficient with a molecular and a turbulent 
contribution.

\subsubsection{Short wavelength approximation}
\label{subsubsec:2-1-2}

The short-wavelength, or WKB, approximation provides a 
unified and comprehensive framework 
for the investigation of the HMRI, AMRI and TI. 
In its simplest version, restricted to the axisymmetric 
HMRI, it traces back to the work 
of Liu et al. \cite{Liu_2006},
which was later corroborated in more detail 
and extended to the non-axisymmetric case  
by Kirillov et al.
\cite{Kirillov_2010,Kirillov_2011,Kirillov_2012,Kirillov_2013,Kirillov_2014}.

Following \cite{Kirillov_2014}, the 
theory starts from the set of equations (1-4)
as described above.
Here we consider the stability of a rotating conducting 
fluid 
exposed to a magnetic 
field sustained by electrical
currents outside and/or inside the fluid. 
Introducing cylindrical coordinates $(r, \phi, z)$ 
we assume a steady-state background liquid flow with          
the angular velocity profile $\Omega(r)$ in a (in general) 
helical 
background magnetic field with constant axial and radially
varying azimuthal components:
\begin{eqnarray}
{\bf{u}}_0(r)=r \,\Omega(r)\,{\bf{
e}}_{\phi},\quad p=p_0(r), \quad {\bf{B}}_0(r)=B_{\phi}^0(r) {\bf{
e}}_{\phi}+B_z^0 {\bf{ e}}_z.
\end{eqnarray}
In the particular case that the azimuthal component is produced 
by an axial current $I$ confined to $r<r_i$, the azimuthal field becomes
\begin{eqnarray}
B_{\phi}^0(r)=\frac{\mu_0 I}{2 \pi r}   .                   
\end{eqnarray}

It is convenient to characterize the shear of the background field 
by the hydrodynamic Rossby number $({\rm Ro})$,
\begin{eqnarray}
{\rm Ro}:=\frac{r}{2\Omega}\partial_r \Omega,
\end{eqnarray}
which gives ${\rm Ro}=0$ for solid body rotation,
${\rm Ro}=-3/4$ for Keplerian rotation, and 
${\rm Ro}=-1$ for the Couette flow 
profile $\Omega(r)\sim r^{-2}$. 
In close correspondence to ${\rm Ro}$, 
we also define the {\it magnetic} Rossby number
\begin{eqnarray}
{\rm Rb}:=\frac{r}{2B_{\phi}^0 r^{-1}}\partial_r (B_{\phi}^0 r^{-1}).
\end{eqnarray}
Then, ${\rm Rb}=0$ results from a linear dependence of
the magnetic field on the radius, $B_{\phi}^0(r)\propto r$, as
it would be produced by a homogeneous axial current in the fluid,
while 
${\rm Rb}=-1$ characterizes the limit of 
a current-free field in the
liquid that is produced solely by an axial current confined to
$r<r_i$.

The linearization around this steady state leads, after some 
algebra, to a
fourth order secular equation for the (complex) spectral 
parameter $\gamma$:
\begin{eqnarray}
p({\gamma})=\det\left( {\bf H}-{\gamma} {\bf I}\right)=0.
\end{eqnarray}
Here,  $\bf I$ is the $4\times4$ identity matrix and $\bf H$ is
\begin{eqnarray}
{\bf H}=\left(
  \begin{array}{cccc}
    -i n {\rm Re} -1 & 2\alpha {\rm Re}  & \frac{i {\rm Ha}(1+n\beta)}{\sqrt{\rm Pm}} & 
    -\frac{2\alpha\beta{\rm Ha}}{\sqrt{\rm Pm}} \\
    -\frac{2{\rm Re}(1+{\rm Ro})}{\alpha} & -i n {\rm Re} -1 & 
    \frac{2\beta{\rm Ha}(1+{\rm Rb})}{\alpha\sqrt{{\rm Pm}}} & 
    \frac{i {\rm Ha}(1+n\beta)}{\sqrt{{\rm Pm}}} \\
    \frac{i {\rm Ha}(1+n\beta)}{\sqrt{\rm Pm}} & 0 & 
    -i n {\rm Re} -\frac{1}{\rm Pm} & 0 \\
    \frac{-2\beta {\rm Ha}{\rm Rb}}{\alpha \sqrt{\rm Pm}} & 
    \frac{i {\rm Ha}(1+n\beta)}{\sqrt{\rm Pm}} & 
    \frac{2{\rm Re}{\rm Ro}}{\alpha} & -i n {\rm Re} -\frac{1}{\rm Pm}\\
  \end{array}
\right)
\label{eqn:operator}
\end{eqnarray}
where we used the magnetic Prandtl number $({\rm Pm})$, 
the ratio of the Alfv\'en frequencies $(\beta)$, the 
Reynolds $({\rm Re})$ 
and Hartmann $({\rm Ha})$ numbers as well as the modified 
azimuthal wavenumber $n$ according to
\begin{eqnarray}
{\rm Pm}=\frac{\omega_{\nu}}{\omega_{\eta}},\quad
\beta=\alpha \frac{\omega_{A_{\phi}}}{\omega_{A_z}},\quad
{\rm Re}=\alpha\frac{\Omega}{\omega_{\nu}},\quad
{{\rm Ha}}=\frac{\omega_{A_z}}{\sqrt{\omega_{\nu}\omega_{\eta}}},\quad
n=\frac{m}{\alpha}.
\end{eqnarray}
Note the definitions for 
the viscous, resistive and the two
$\rm Alfv\acute{e}n$ frequencies:
\begin{eqnarray}
\omega_{\nu}={\nu} |{\bf{ k}}|^2,\quad \omega_{\eta}={\eta}|{\bf{ k}}|^2,
\quad {\omega_{A_z}=\frac{k_z {B_z^0}}{\sqrt{\rho \mu_0}}},
\quad \omega_{A_{\phi}}=\frac{B_{\phi}^0}{R \sqrt{\rho\mu_0}},
\end{eqnarray}
with ${\bf{k}}=(k^2_z+k^2_r)^{1/2}$ and $\alpha=k_z/|{\bf{k}}|$.

\begin{figure}[h]
\includegraphics[width=0.9\textwidth]{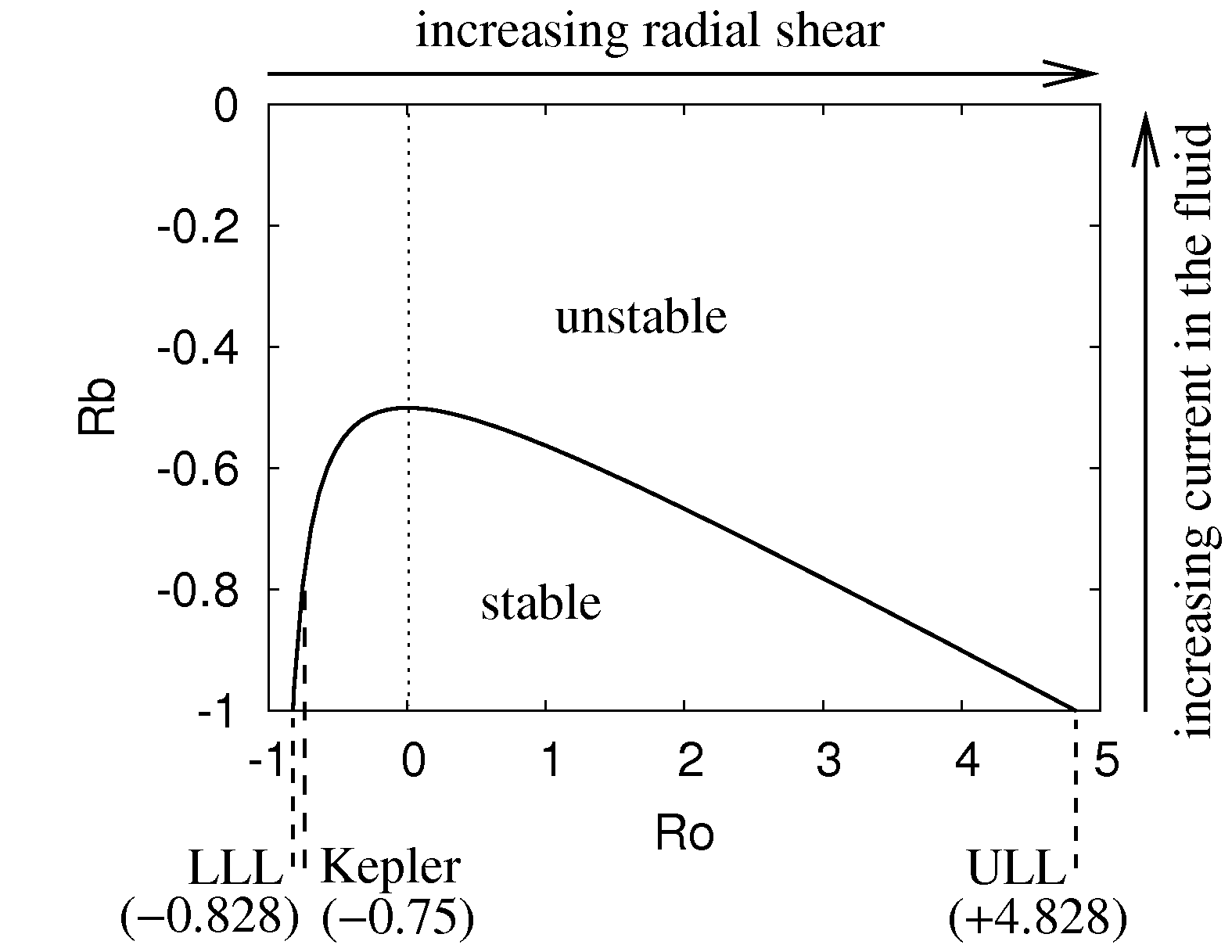}
\caption{Stability chart in the ${\rm Ro}-{\rm Rb}$ plane, 
for ${\rm Pm}=0$, ${\rm Ha}$ and ${\rm Re}$ tending to infinity, 
and optimized values of $\beta$ (for HMRI) or $\alpha$ (for AMRI). 
The Liu limits LLL and ULL apply only for 
${\rm Rb}=-1$, while for ${\rm Rb}>-1$ 
shallower shear profiles can also be destabilized (including Kepler
rotation starting at ${\rm Rb}=0.78125$.). The dotted line separates flows
with negative shear (to the left) and positive shear (to the right).}
\label{fig:ro_rb}    
\end{figure}

The secular equation (\ref{eqn:operator}) has been analyzed in much detail 
using Bilharz' stability criterion \cite{Kirillov_2013,Kirillov_2014}. One 
of the most important 
results refers to the limiting stability curve 
(see Fig.~\ref{fig:ro_rb}) 
in the ${\rm Ro}-{\rm Rb}$ plane, which obeys the analytical equation
\begin{eqnarray}
{\rm Rb}=-\frac{1}{8}\frac{({\rm Ro}+2)^2}{{\rm Ro}+1} \;.
\end{eqnarray}
The crossing points of this curve with the abscissa 
(${\rm Rb}= -1$) recover the original result of \cite{Liu_2006},
that for ${\rm Pm}=0$, and optimal values of $\beta$, 
the instability domains lie outside the stratum
$$
2-2\sqrt{2}=:{\rm Ro}_{\rm LLL}<
{\rm Ro}<{\rm Ro}_{\rm ULL}:=2+2\sqrt{2},
$$
where ${\rm Ro}_{\rm LLL}=2 (1-\sqrt{2}\approx -0.828$ denotes the
lower Liu limit (LLL), as we call it now, and
${\rm Ro}_{\rm ULL}= 2(1+\sqrt{2})\approx 4.828$ 
the upper Liu limit (ULL).

Figure~\ref{fig:ro_rb} shows that the LLL and ULL  are just the 
endpoints of a quasi-hyperbolic curve. At ${\rm Rb}=-1/2$ 
the branches for negative shear ${\rm Ro}<0$ and positive shear 
${\rm Ro}>0$ meet each other. 
We find that in the inductionless case ${\rm Pm}=0$, when 
the Reynolds and 
Hartmann numbers  tend to infinity in a well defined 
manner \cite{Kirillov_2014}
and $\beta$ or $n$ are optimized, the maximum 
achievable critical Rossby number ${\rm Ro}_{\rm extr}$ 
increases with the  increase of ${\rm Rb}$. At
${\rm Rb}\ge -25/32=-0.78125$, 
${\rm Ro}_{\rm extr}$  would even exceed the critical 
value for the Keplerian flow.
Therefore, the very possibility for $B_{\phi}(r)$ to depart 
from the profile $B_{\phi}(r)\propto r^{-1}$  allows us to break 
the conventional lower Liu limit and extend the inductionless versions 
of MRI to velocity profiles $\Omega(r)$ as flat as the Keplerian 
one and even to less steep profiles.

One might ask whether there is any deeper physical 
sense behind the two Liu limits and the line connecting them.
In order to clarify this point, 
we have investigated in \cite{Mama_2016} the non-modal 
dynamics of HMRI arising from the non-normality of the operator
$\bf H$ in Eq.~(\ref{eqn:operator}).
For $\rm Pm \ll 1$ we traced the
entire time evolution of the optimal growth of the 
perturbation energy and demonstrated how the non-modal
growth smoothly carries over to the modal 
behavior at large times.  At small and intermediate 
times, HMRI undergoes a transient amplification.
It turned out that the modal
growth rate of HMRI displays a very similar dependence
on $\rm Ro$ as the maximum non-modal growth of the purely
hydrodynamic shear flow.
In particular, the optimized non-modal growth rate 
$G_m$ turned out to be {\it identical} at the two Liu limits,
namely $G_m=(1+{\rm Ro})^{sign({\rm Ro})}= 5.828$!
This indicates a fundamental
connection between non-modal dynamics and dissipation-
induced modal instabilities, such as HMRI. Despite
the latter being magnetically triggered, both 
rely on hydrodynamic means of amplification, i.\,e., they 
extract energy from the background flow mainly by Reynolds 
stress.

\subsubsection{1D and 3D simulations}

Actually, the apparently simpler WKB analysis of HMRI, 
AMRI and TI, as sketched above, was 
preceded by many detailed 1D and 3D simulations
\cite{Hollerbach_2005,Hollerbach_2010,Ruediger_2005,Ruediger_2007}.
Specifically, the first predictions of 
HMRI \cite{Hollerbach_2005,Ruediger_2005} and AMRI 
\cite{Hollerbach_2010,Ruediger_2007} were done with 1D 
codes solving a linear eigenvalue problem for 
determining the stability thresholds.

\begin{figure}[h]
\includegraphics[width=0.95\textwidth]{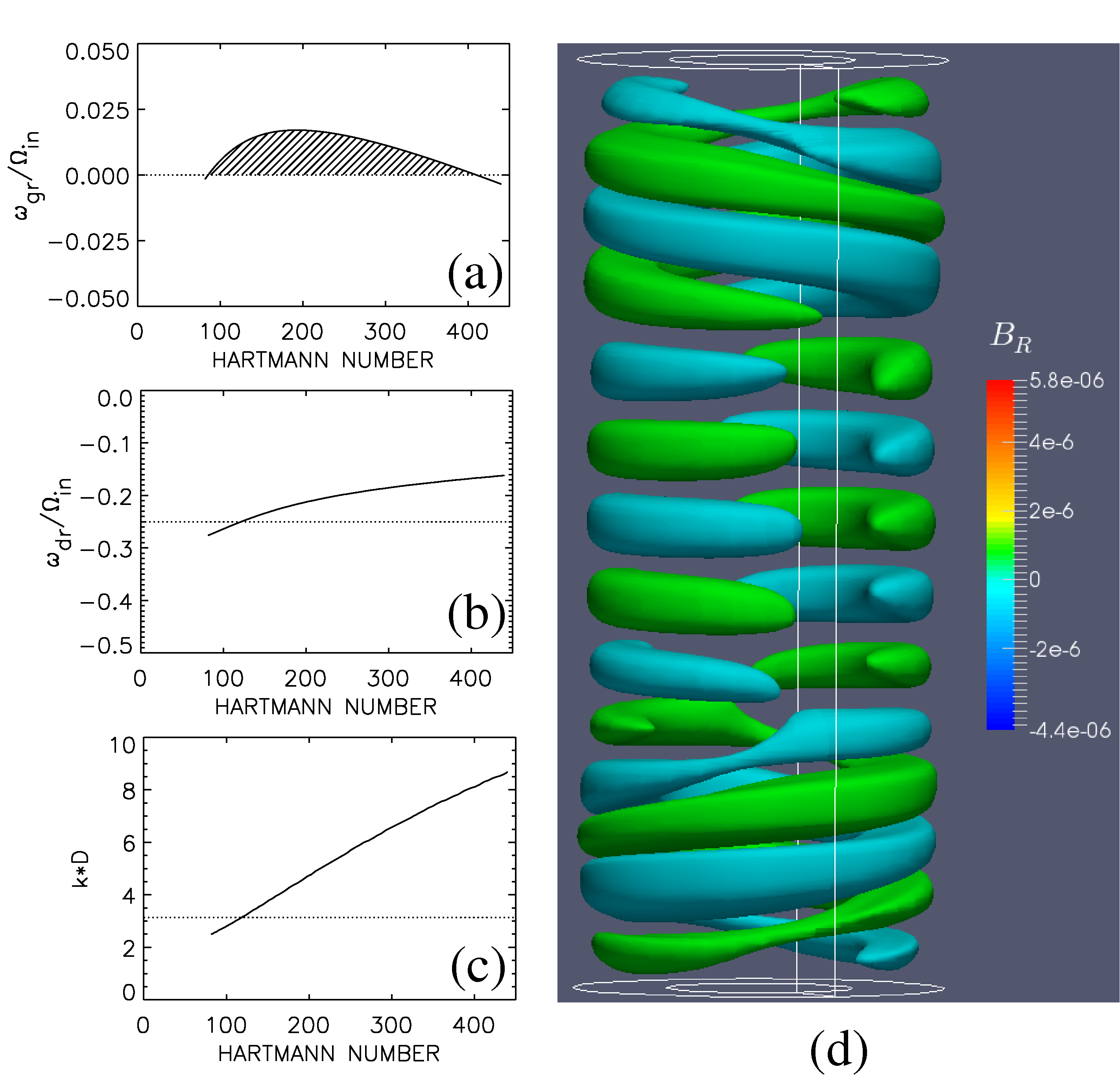}
\caption{Numerical simulations for the AMRI experiment. 
Left: Results of a 1D linear eigenvalue solver
giving the normalized growth rate (a), the normalized 
drift frequency (b), and the normalized wave number (c).
(d) Radial magnetic field
computed with a 3D nonlinear code, at $\rm Re=1500$, 
$\rm Ha=100$, $\mu=0.26$ and $\rm Pm=10^{-5}$ with split end-plates 
in units of the applied external field.}
\label{fig:simulation}      
\end{figure}

Figure ~\ref{fig:simulation} shows a recent example
for predicting the AMRI experiment at the PROMISE
facility (to be discussed below). For the experimentally relevant 
values ${\rm Pm}=10^{-6}$, ${\rm Re}=3000$,  
$\mu:=f_o/f_i=0.26$, and 
varying $\rm Ha$, the left column shows 
the normalized growth rate (a), the normalized 
drift rate (b), and the normalized wave number (c), all
obtained with the linear eigenvalue solver \cite{Ruediger_2014}.
Evidently, the unstable region extends  
from ${\rm Ha}\approx 80$ to 
${\rm Ha}\approx 400$.
The right column (d) shows the radial magnetic field
component of the AMRI 
pattern which was computed with the 3D fully 
non-linear code described 
in \cite{Gellert_2007}.

More recent numerical results include the 
proof that Chandrasekhar's supposedly stable solution 
\cite{Chandra_1956}, in which the Alfv\'en velocity of the 
azimuthal field equals the rotation velocity of the background 
flow, becomes unstable against non-axisymmetric perturbations 
if at least one of the diffusivities is finite 
\cite{Ruediger_2015}. 
Another example of such (double-)diffusive instabilities
is the magnetic destabilization of flows with 
positive shear \cite{Ruediger_2016}.

\subsubsection{Turbulent diffusion}

Observations of the present sun with its
nearly rigidly rotating core, and
of far-developed main-sequence stars,
lead to the conclusion 
that a mechanism must exist that transports angular 
momentum rather effectively outwards. 
Since the radiative 
cores are stably stratified, 
convection is most probably not relevant
here. Furthermore, 
the sought-after mechanism should  enhance the transport of 
chemicals only
mildly, with an effective diffusion coefficient 
$D_{\rm turb}$ exceeding 
the molecular viscosity by one or two orders 
of magnitude \cite{Schatzmann_1977}. The authors of 
\cite{Lebreton_1987} considered a 
relation $D_{\rm turb}={\rm Re}^* \nu$ 
for the diffusion coefficient (after the notation of Schatzman, 
see \cite{Zahn_1990}) with ${\rm Re}^*\simeq 100$. 
While it has been known for some time that
magnetic instability could provide enhanced angular 
momentum transport (see \cite{Spada_2016}), we
show in the following that it also leads to a Schatzman number 
${\rm Re}^*$ of the correct magnitude.

The numerical model is based on a Taylor-Couette set-up 
with a current-driven magnetic field that becomes unstable 
due to the Tayler instability.
The whole system rotates differentially with a quasi-Keplerian rotation 
profile ($\mu=0.354$ for a radius ratio of 
$r_i/r_o=0.5$). The most unstable mode is the $m=1$ that 
grows and saturates. Now the additional diffusion 
equation (\ref{eqn:eq_tdiff}) is 
switched on and the enhanced or turbulent diffusion coefficient 
$D_{\rm turb}$ due to the TI is measured 
(in units of the molecular diffusion coefficient 
$D_{\rm mol}$) \cite{Paredes_2016}.

\begin{figure}[h]
\includegraphics[width=0.99\textwidth]{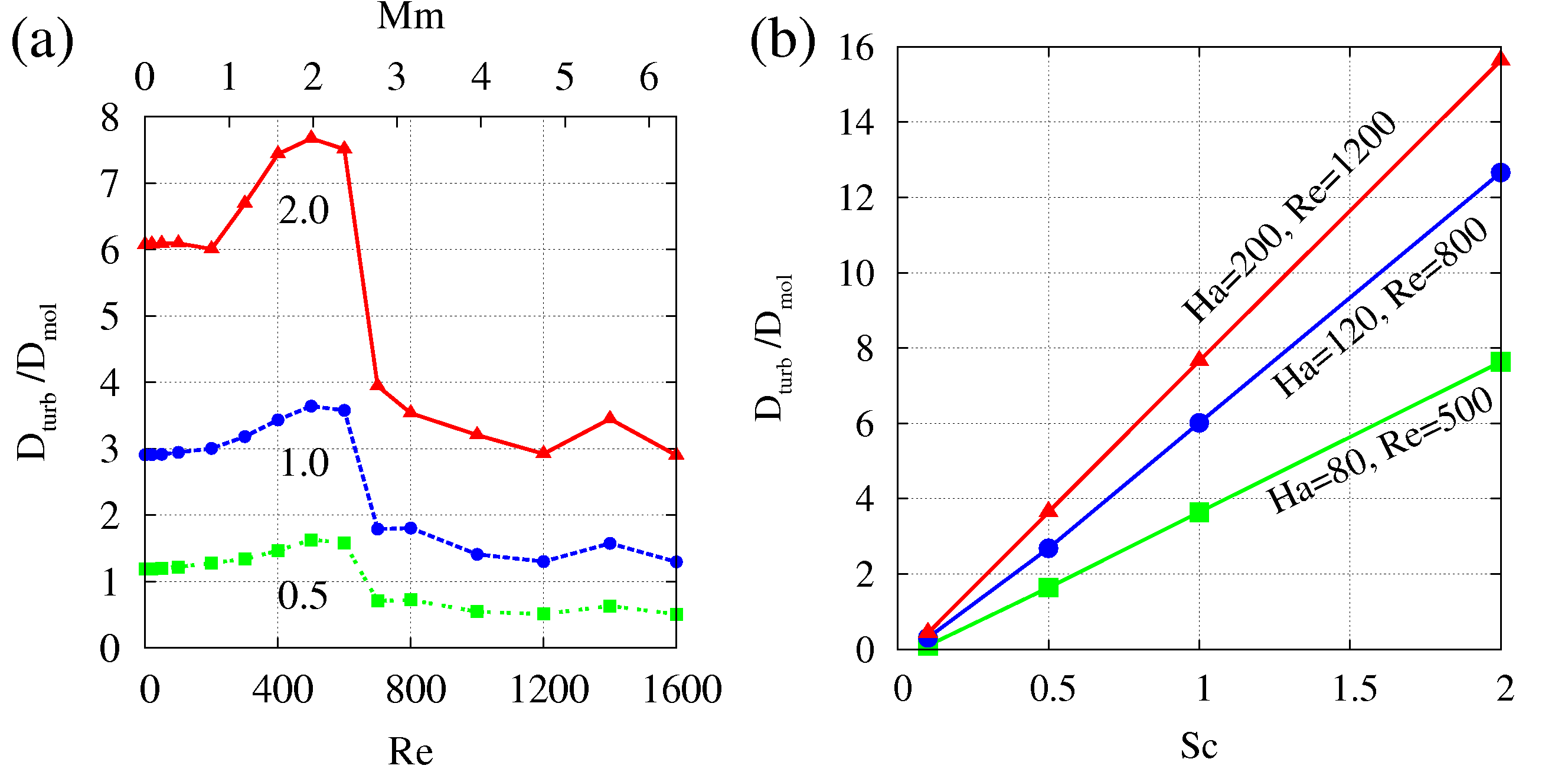}
\caption{Turbulent contribution to the diffusivity or mixing of 
 TI-unstable flows. 
         (a) Reynolds number dependence of $D_{\rm turb}$ for 
	 three values of the Schmidt number ${\rm Sc}=0.1/1/2$ 
	 with ${\rm Ha}=80, {\rm Pm}=0.1$. The value 
         ${\rm Mm}=1$ (top horizontal axis) separates the regimes 
	 of slow and fast rotation. Note the reduction of 
	 $D_{\rm turb}/D_{\rm mol}$ for ${\rm Mm}>2$.
         (b) Schmidt number dependence with increasing 
	 slope or Schatzman ${\rm Re}^*$. Always $\mu=0.35$.}
\label{fig:tdiff}
\end{figure}

The resulting ratio $D_{\rm turb}/D_{\rm mol}$ for a fixed Hartmann 
number is shown in Fig.~\ref{fig:tdiff}a.
In the slow rotation regime (with a magnetic Mach number 
${\rm Mm} = {\rm Rm}/{\rm S} <1$), the effective diffusivity hardly 
changes and is very close to zero. In the fast rotation regime, the 
ratio $D_{\rm turb}/D_{\rm mol}$ increases monotonically until 
it reaches a maximum at about ${\rm Mm}\simeq 2$. Finally, for faster 
rotation (${\rm Mm}>2$), the effective diffusivity decreases rapidly 
due to the rotational quenching and reaches a rather constant value. 
Nevertheless, the effective normalized diffusivity always 
increases with increasing Schmidt numbers so that the Schatzman 
relation $D_{\rm turb}\propto \nu$ is valid without any influence 
from the microscopic diffusivity.

Figure ~\ref{fig:tdiff}b shows the normalized 
diffusivity for fixed Hartmann number as a function
of ${\rm Sc}$. For all Reynolds numbers, the induced diffusivities 
$D_{\rm turb}$ are different from zero, and the 
ratio $D_{\rm turb}/D_{\rm mol}$ scales linearly with ${\rm Sc}$. 
The figure only shows the relation for those 
Reynolds numbers for which the diffusivities reach their maximum 
at ${\rm Mm}\approx2$. For  ${\rm Sc}\to 0,$ the 
diffusivity $D_{\rm turb}$ vanishes as well. However, the essential 
result is the linear relation between 
$D_{\rm turb}/D_{\rm mol}$ and ${\rm Sc}$ for molecular Schmidt 
numbers ${\rm Sc}>0.1$. In the notation of Schatzman 
this means $ D_{\rm turb} = {\rm Re}^* \ \nu$ with the scaling 
factor ${\rm Re}^*$ (which indeed forms  some kind 
of Reynolds number).  This linear relation holds for all 
considered Reynolds and Hartmann numbers.

The scaling or {\em Schatzman} factor ${\rm Re}^*$  increases 
with increasing ${\rm Ha}$, while for all three 
shown parameter sets the magnetic Mach number is nearly the 
same. We find ${\rm Re}^* \sim 4$ for ${\rm Ha}=80$ 
increasing to ${\rm Re}^* \sim 8$ for ${\rm Ha}=200$. A 
saturation for larger ${\rm Ha}$ is indicated by the 
results presented in Fig. \ref{fig:tdiff} and would be of 
the order of 50, which is indeed 
the right magnitude of the additional
diffusion process acting in the sun.

\subsection{Experiments on HMRI}
\label{subsec:2-2}

\begin{figure}[h]
\includegraphics[width=0.99\textwidth]{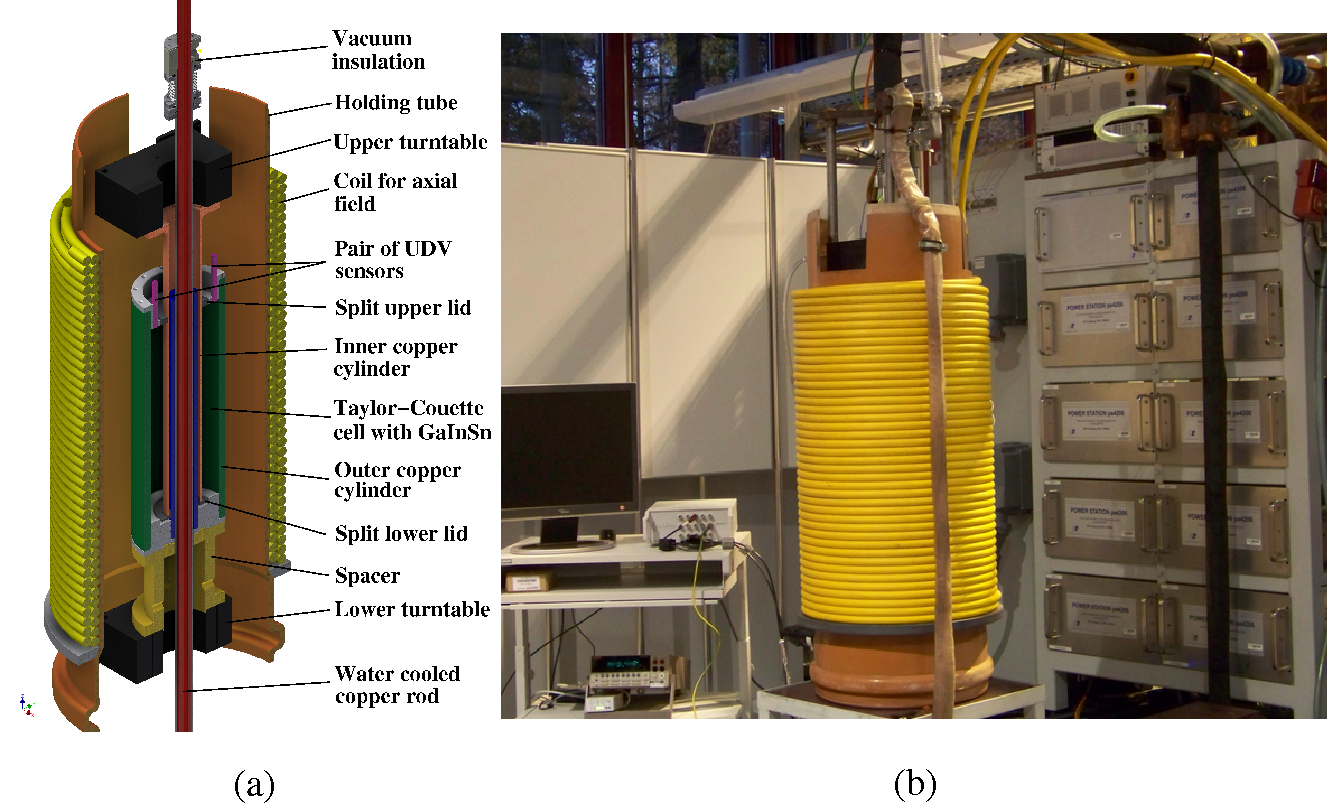}
\caption{The PROMISE experiment: (a) Schematic sketch of the central
Taylor-Couette set-up and the coil for producing the axial field.
(b) Photograph of the installation, with the 20\,kA power supply 
visible on the right.}
\label{fig:anlage}       
\end{figure}

Shortly after the theoretical prediction of HMRI 
\cite{Hollerbach_2005}
the PROMISE experiment was set-up at HZDR. Its heart 
is a cylindrical vessel made of copper (Fig.~\ref{fig:anlage}).
The inner wall extends in radius from 22 to 32 mm; the outer
wall extends from 80 to 95 mm.
This vessel is filled with the alloy GaInSn, which is 
a most convenient medium 
as it is liquid at room temperatures. The physical properties
of GaInSn at 25$^{\circ}$C are: density $\rho=6.36 \times 10^3$ kg/m$^3$,
kinematic viscosity $\nu=3.40 \times 10^{-7}$m$^2$/s, electrical 
conductivity $\sigma= 3.27 \times 10^6$ S/m. The magnetic Prandtl
$\rm Pm$
is therefore $1.4 \times 10^{-6}$.

The copper vessel is fixed via a spacer 
on a precision turntable. Hence, the outer wall of the vessel
serves also as the outer cylinder of the Taylor-Couette flow. The inner 
cylinder of the Tayler-Couette flow is fixed to an upper turntable,
and is immersed into the GaInSn from above. It has a
thickness of 4 mm, extending from 36 to 40 mm.
The actual Taylor-Couette flow then extends between $r_i= 40$ mm
and $r_o=80$ mm. In the very first set-up, PROMISE 1 
\cite{Stefani_2006,Stefani_2007}, the upper end-plate
was a plexiglass lid fixed to the frame while the
bottom was simply part of the copper vessel, and hence
rotated with the outer cylinder.  There was thus a clear
asymmetry in the end-plates, with respect to both their
rotation rates and electrical conductivity.

In the second version, PROMISE 2 \cite{Stefani_2009},
the lid-configuration was significantly improved by
using insulating rings both on top and bottom, and
by splitting them at a well defined intermediate radius 
of 56 mm which had been found in \cite{Szklarski_2007} 
to minimize the Ekman pumping.

\begin{figure}[h]
\includegraphics[width=0.99\textwidth]{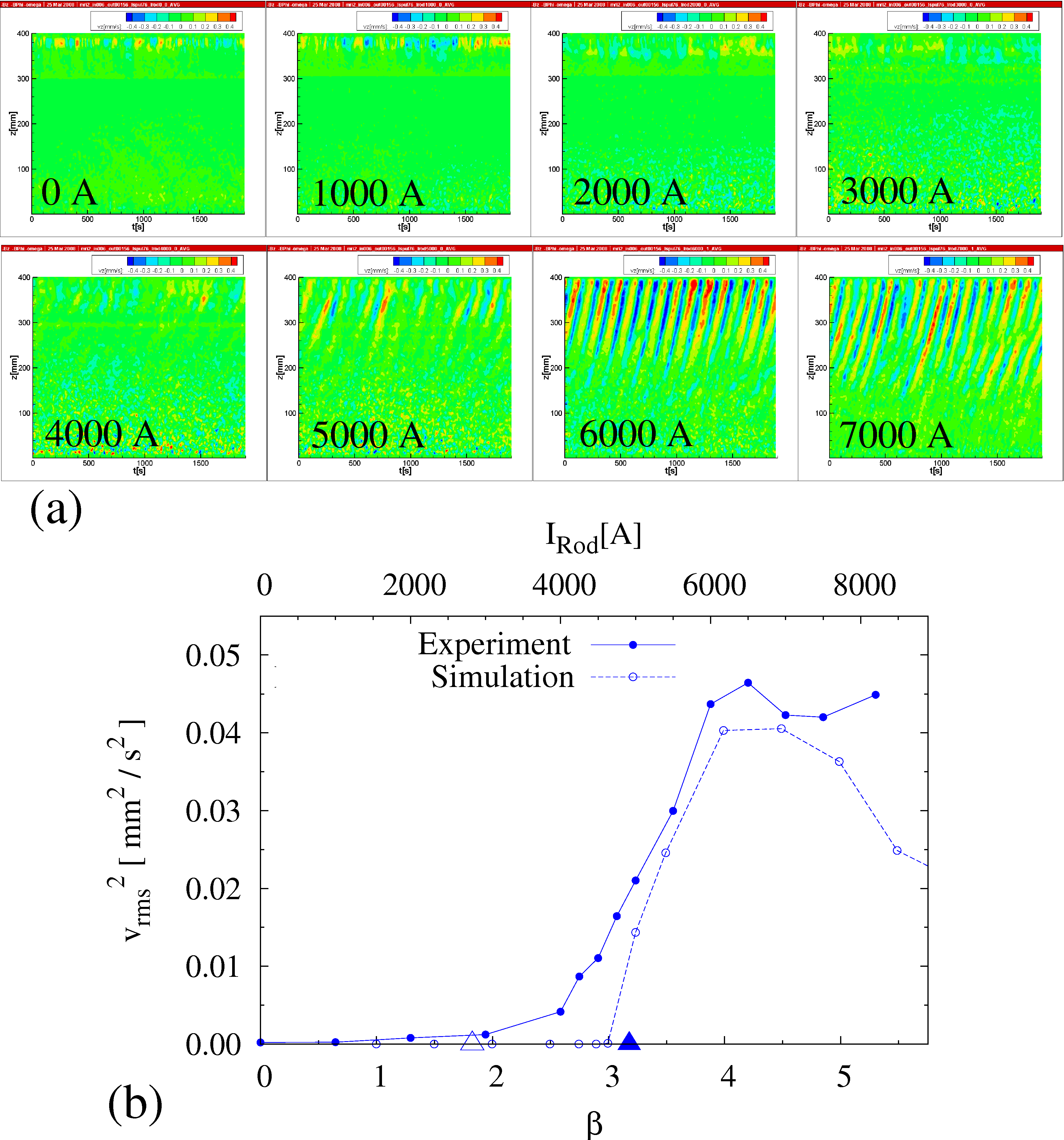}
\caption{Results of the PROMISE experiment, when varying the axial
current. The fixed parameters are $f_i=0.06$\,Hz, $f_o=0.0156$\,Hz, 
resulting 
in $\rm Re=1775$ and $\mu=0.26$, $I_{\rm coil}=76$\,A, giving 
$\rm Ha=12$. (a) Measured UDV signals in dependence on time  (abscissa) 
and vertical position (ordinate axis), for 8 different axial
currents $I_{\rm rod}$. (b) Rms of the 
perturbation of the axial velocity
at the UDV sensor position, in dependence on $I_{\rm rod}$. The dashed
line shows the numerical predictions, the full line shows
the experimental results. The triangles give the numerical
prediction for the onset of the convective instability (empty triangle)
and the absolute instability (full triangle).
}
\label{fig:betalinie1}      
\end{figure}

With this set-up, HMRI has been comprehensively
characterized by varying various parameters and 
comparing the observed  travelling wave structure with  
numerical predictions. The variations included those
of the Reynolds number $\rm Re$, of the ratio 
$\mu=f_o/f_i$ of rotation rates, 
of the Hartmann number $\rm Ha$, and
of the ratio $\beta$ of azimuthal to axial field.
Figure~\ref{fig:betalinie1} illustrates a typical result for the 
latter variation. Fixing $f_i=0.06$\,Hz, $\mu=0.26$,
$I_{\rm coil}=76$\,A, we varied the axial current between 0 
and 8.2\,kA. We observe a travelling HMRI wave 
only  above 4\,kA (Fig.~\ref{fig:betalinie1}a). 
The comparison 
with numerical predictions
is shown in Fig.~\ref{fig:betalinie1}b.
The two triangles indicate the numerically determined thresholds
for the convective and the absolute instability. 
The experimental curve suggests that the observed instability 
is indeed a global one.

\subsection{Experiments on AMRI}
\label{subsec:2-3}
What happens in the PROMISE experiment when we switch 
off the axial magnetic field? The theoretical predictions 
for this case have been given in \cite{Hollerbach_2010}. 
The axisymmetric ($m=0$) HMRI is then replaced by the
non-axisymmetric ($m=1$) AMRI. At the same time, the threshold 
of the Hartmann number increases to 80 which requires 
a central current of approximately 10\,kA 
\cite{Seilmayer_2014}. 

In order to explore that parameter region, the 
power supply for the central rod, restricted 
previously to 8 kA, was replaced by a new
one which is able to deliver 20\,kA.
Connected with this enhancement of the switching 
mode power supply, significant effort had to be spent 
on various issues of electromagnetic interference 
\cite{Seilmayer_2016}, 
before the (initially extremely noisy) 
UDV data could be utilized for characterizing the 
AMRI.

\begin{figure}[h]
\includegraphics[width=0.99\textwidth]{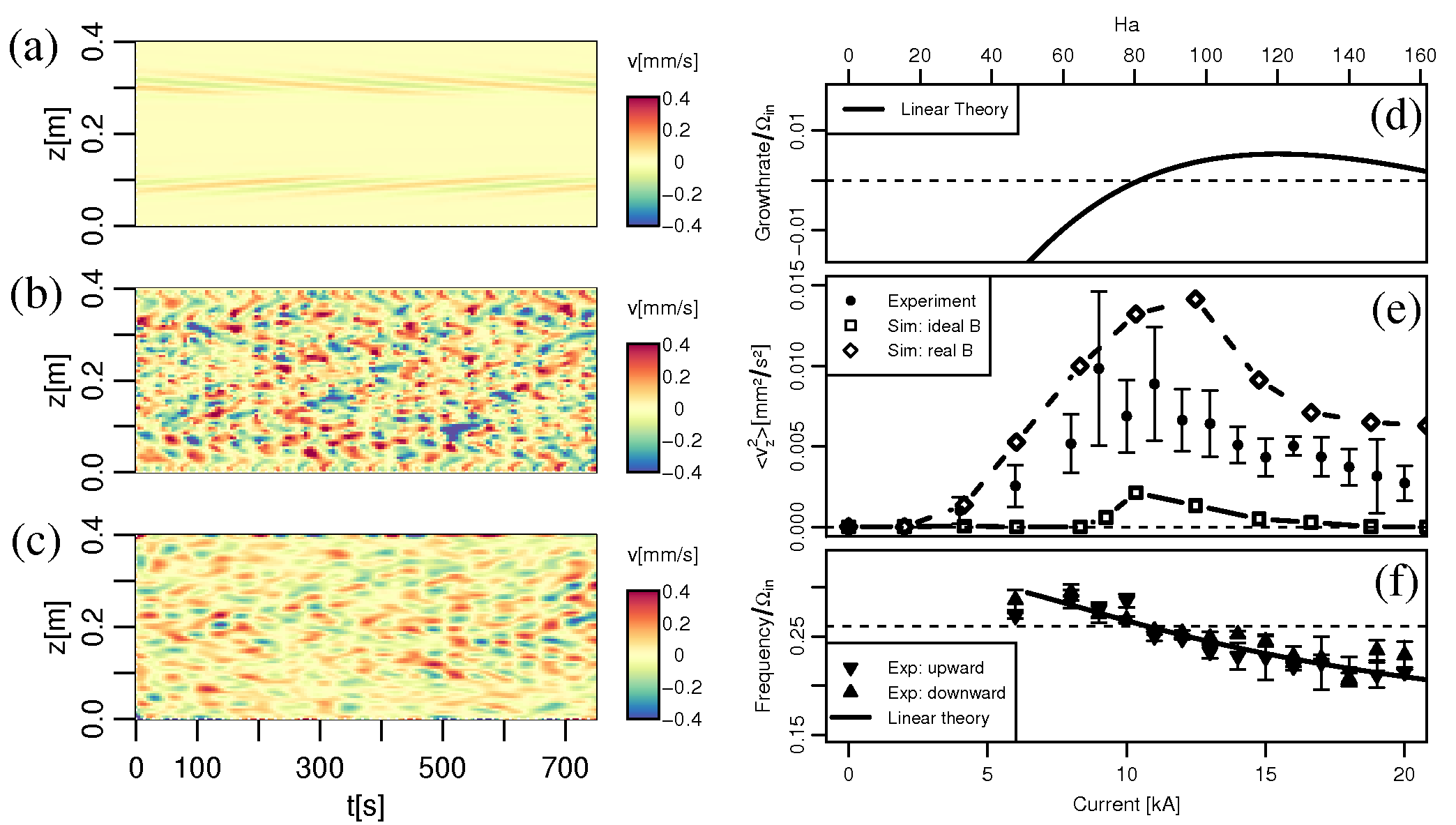}
\caption{Results of the AMRI experiment. Left:
Velocity perturbation $v_z(m=1,z,t)$ for $\mu=0.26$, $\rm Re=1480$,
and ${\rm Ha}=124$. (a) Simulation for ideal axisymmetric field. 
(b) Simulation for realistic field. (c) Experimental results. Right:
Dependence of various quantities on $\rm Ha$: 
(d) Numerically determined growth rate. (e) Mean 
squared velocity perturbation. (f) Angular drift frequency. 
In the frequency plot, ``upward'' and ``downward'' 
refer to the travel direction of the AMRI wave. 
After \cite{Seilmayer_2014}.} 
\label{fig:amri_ergeb}
\end{figure}

Another problem that became evident when analyzing the data 
was the
factual slight asymmetry of the allegedly axisymmetric 
field which results
from the one-sided wiring of the central vertical current
(Fig.~\ref{fig:anlage}b).
While the observable AMRI structure was originally expected to 
be similar to that shown in Fig.~\ref{fig:simulation}d,
a more sophisticated numerical analysis 
revealed the specific effect of the slight symmetry breaking.
This is illustrated in Fig.~\ref{fig:amri_ergeb}a which 
shows the spatio-temporal structure
of the AMRI wave as it would be expected for a perfectly symmetric,
purely azimuthal magnetic field produced by infinitely long wire 
at the central axis. The parameters for this simulations are 
$\rm Re=1479$,  $\rm Ha=124$.
What comes out here is a very regular ''butterfly'' diagram of
downward and upward traveling waves which are concentrated in the
upper and lower half of the cylinder, respectively. 
While in an infinitely long cylinder
upward and downward travelling waves should be equally likely, 
it is the breaking of axial symmetry due to the effect of the end-caps
that induces a preference for upward or downward travelling waves 
in the two halves of the vessel. 

Interestingly, the effect of this first symmetry breaking 
in axial
direction (due to the end-caps) is neutralized by the second 
symmetry breaking in azimuthal direction (due to the one sided 
wiring). As a result, the upward and downward travelling waves 
interpenetrate each other in the two halves of the cylinder. 
This effect has been revealed both in a numerical
simulation which respects the correct geometry 
(Fig.~\ref{fig:amri_ergeb}b), as 
well as in the experiment (Fig.~\ref{fig:amri_ergeb}c).
In Fig.~\ref{fig:amri_ergeb}d-e we summarize the results for 
varying Hartmann numbers, including the simulated growth rate
(d), 
the measured and predicted rms of the velocity perturbation (e), 
and the measured and simulated drift frequencies (f).

While the observed and numerically confirmed effects of the
double symmetry breaking on the AMRI are interesting in their own
right, we have decided to improve the experiment in such a way 
that the azimuthal symmetry breaking is largely avoided.
This is accomplished by a new system of wiring of the central current, 
comprising now a ''pentagon'' of 5 back-wires situated around 
the experiment. 
First experiments with this set-up show encouraging results.
Any details, in particular on the transitions between
AMRI and HMRI  when cranking up the axial magnetic field, are 
left for future publications.

\subsection{Experiments on TI}
\label{subsec:2-4}

Imagine the central current used in the AMRI experiment to be
complemented, or completely replaced, by a parallel current guided
through the fluid column. This current provides a further 
energy source for instabilities that adds to any 
prevailing differential rotation. In the limiting case 
of vanishing differential rotation, the purely
current-driven, kink-type Tayler instability (TI) 
will arrive for sufficiently large currents.

In principle, this effect has been known for a long time
in plasma physics, where the (compressible and non-dissipative) 
counterpart of TI is better known as the kink instability 
in a $z$-pinch \cite{Bergerson_2006}.
In astrophysics, TI has been discussed as 
a possible ingredient of an alternative, nonlinear stellar 
dynamo mechanism (Tayler-Spruit dynamo \cite{Spruit_2002, Stefani_2016a}), 
as a mechanism for generating helicity \cite{Gellert_2008}, 
and as a possible source of helical structures in galactic jets 
and outflows \cite{Moll_2008}. 
A particular, though non-astrophysical, motivation to study 
the TI in liquid metals arises from the growing interest 
in large-scale liquid metal batteries
as supposedly cheap  storages for 
renewable energies. 
Such a battery would consist of a self-assembling stratification 
of a heavy liquid half-metal (e.\,g. Bi, Sb) at the bottom, an 
appropriate molten salt as electrolyte in the middle, 
and a light alkaline or earth alkaline metal (e.\,g. Na, Mg) at the top. 
While small versions of this battery have already been tested 
\cite{Kim_2012}, for larger versions the occurrence of TI could represent a 
serious problem for the integrity of the stratification
\cite{Stefani_2011,Stefani_2016,Weber_2013,Weber_2014,Weber_2015}. 

For liquid metals, TI is expected to set in at
an electrical current in the order of kA.
The precise value is a function of  
various material parameters,
since for viscous and resistive fluids the TI is 
known \cite{Montgomery_1993,Ruediger_2011,Spies_1988} 
to depend effectively on the Hartmann number 
${\rm Ha}=B_{\varphi}(R) R (\sigma/(\rho \nu)^{1/2}$, where $R$ is 
the radius of the column. 

Here, we summarize experimental results \cite{Seilmayer_2012}
that confirm the numerically determined growth rates of TI 
\cite{Ruediger_2011,Stefani_2011} as well as the
corresponding prediction  that the critical
current increases monotonically with the 
radius of an inner cylinder. 

\begin{figure}[h]
\includegraphics[width=0.95\textwidth]{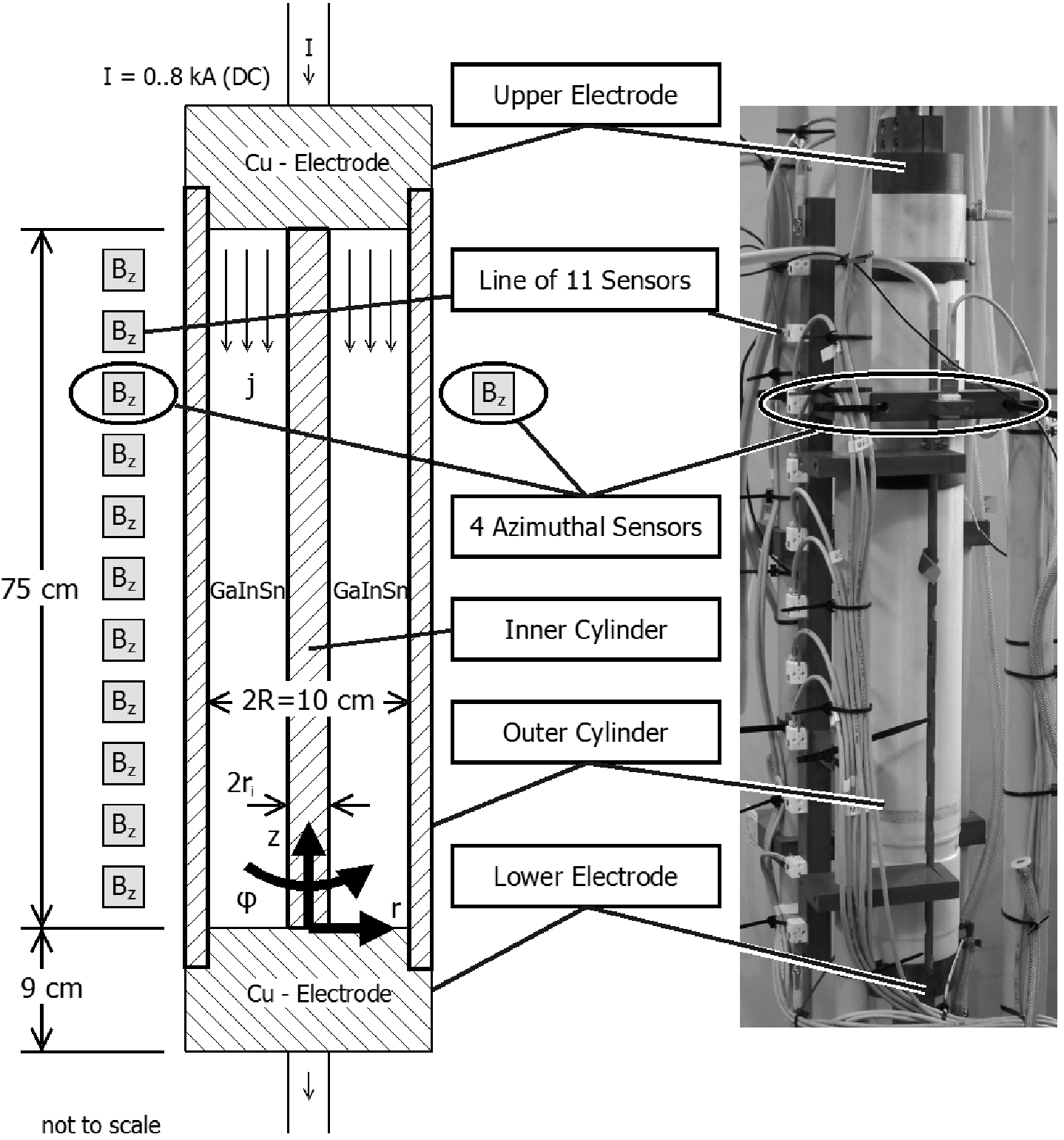}
\caption{Set-up of the TI-experiment. Left: Scheme 
with liquid metal column and fluxgate sensors
positioned along the vertical axis and the azimuth. 
Right: Photograph of the central part of the experiment.}
\label{fig:ti_aufbau}       
\end{figure}

The central part of our TI experiment (Fig.~\ref{fig:ti_aufbau}) 
is an insulating cylinder 
of length 75\,cm and inner diameter 10\,cm, filled again 
with the eutectic alloy GaInSn. 
At the top and bottom, the liquid metal column is contacted 
by two massive copper electrodes of height 9\,cm which are connected 
by water cooled copper tubes to a DC power supply that is able to 
provide electrical currents of up to 8\,kA. By intensively rubbing the 
GaInSn into the copper, we have provided a good wetting
making the electrical contact 
as homogeneous as possible.
 
For the identification of TI we exclusively relied on the signals of 
14 external fluxgate sensors that measure the vertical 
component $B_z$ of the magnetic field. Eleven of 
these sensors were aligned along the vertical axis 
(with a spacing of 7.5\,cm), while the remaining three sensors were 
positioned along the azimuth in the upper part, approximately 
at 15\,cm from the top electrode. 
The distance of the sensors from the outer rim of the
liquid metal column was 7.5\,cm. This rather large value, which
is certainly not ideal to identify small wavelength perturbations, 
has been chosen in order to prevent saturation of the fluxgate 
sensors in the strong azimuthal field of the axial current.

One of the main goal of the experiment was to study the influence 
of the electric current through the fluid on the growth rate and 
on the amplitude of the magnetic field perturbations. 
This was done without any insert, as well as 
for two different radii of an inner non-conducting cylinder, 
$r_i$=6\,mm and 12.5\,mm, for which we expect a monotonic increase of 
the critical current with the radius.

\begin{figure}[h]
\includegraphics[width=0.8\textwidth]{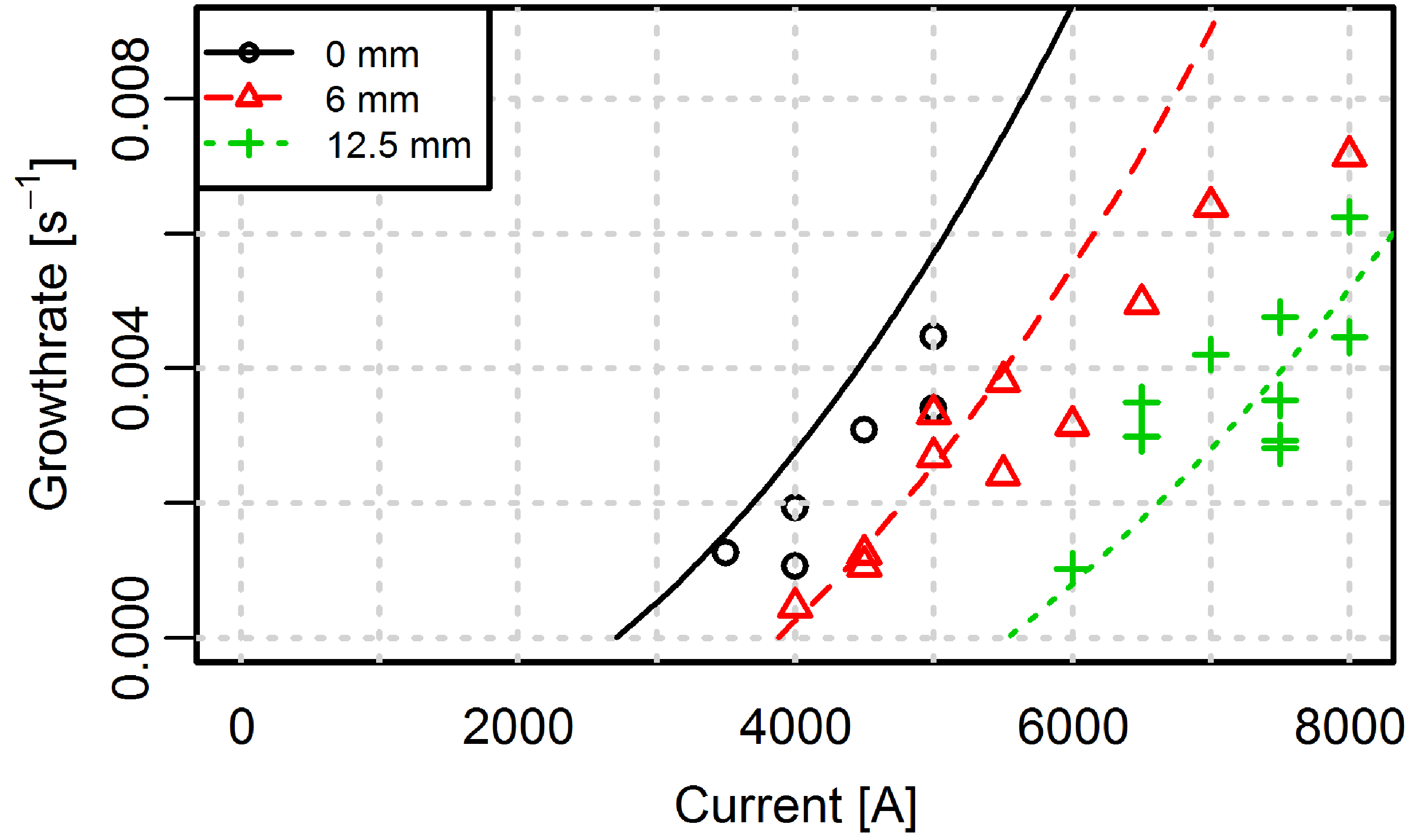}
\caption{Growth rates of the TI in dependence on the current 
for three different radii (0, 6, and 12.5 mm) 
of a central insulating cylinder. The lines give the numerical results,
the symbols show the values inferred from the experiment. After
\cite{Seilmayer_2012}.}
\label{fig:lambda_ti}      
\end{figure}

In Fig.~\ref{fig:lambda_ti} we compile the dependence of 
the growth rate of the TI 
on the radius of the inner cylinder and 
on the current through the fluid, and compare them with 
the numerically determined growth rates 
\cite{Ruediger_2011} for the mode with optimum 
wavelength (appr. 13 cm).  Despite some scatter of the 
data, we observe a quite reasonable agreement with the numerical 
predictions.

We have to admit that the realization and the data 
analysis of the TI experiment was significantly harder
than originally envisioned. The main reason is the indirect 
identification of the TI by the fluxgate sensors along the 
height of the column and the azimuth. 
The very weak induced magnetic fields, the slightly
under-critical number of sensors, but also the 
Joule heating due to the high current densities 
were main obstacles for a clear identification
of the TI.

We note that some later experiments using 
Ultrasonic Doppler Velocimetry (UDV) in order to measure
directly the axial velocity component along the $z$-axis
did not really solve the problems \cite{Starace_2015}. Actually,
they  even lead to strong electro-vortex flows driven by the
inhomogeneous current distribution at the sensor positions
which spoiled significantly the previous TI results 
obtained with a
more homogeneous interface between fluid and 
copper electrodes.

\section{Instabilities in spherical geometry}
\label{sec:3}

\begin{figure}[h]
\includegraphics[width=0.95\textwidth]{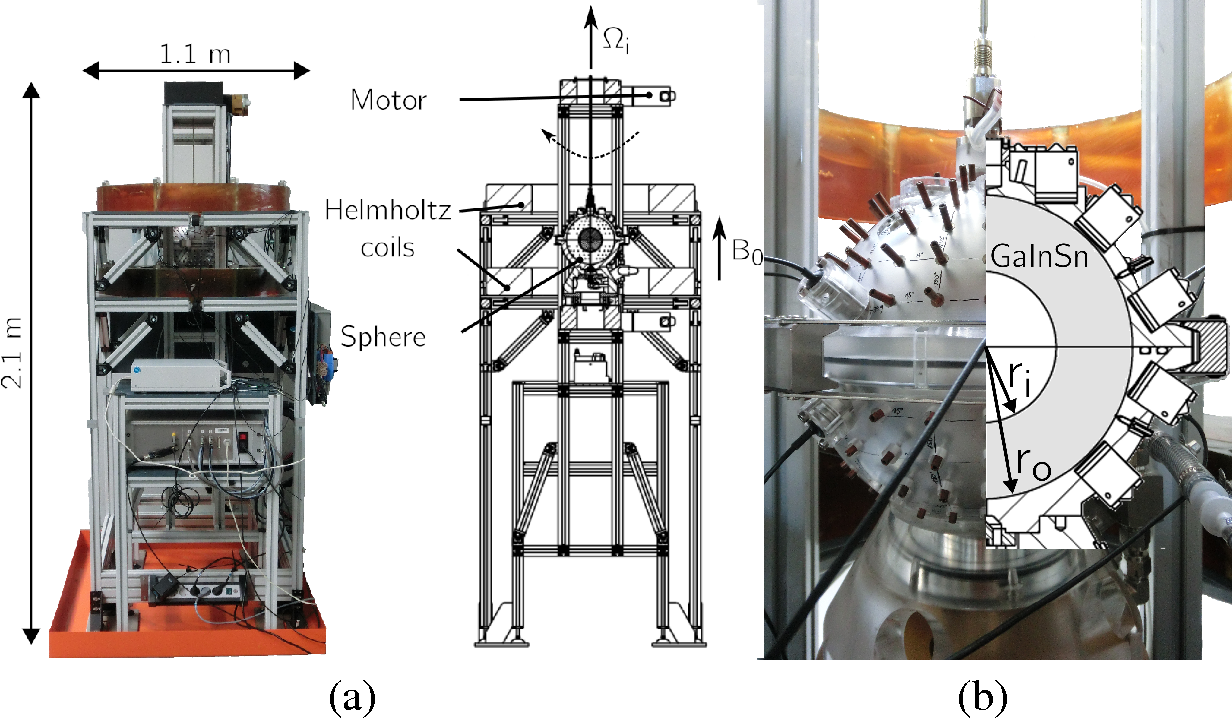}
\caption{The HEDGEHOG experiment: (a) Photograph and drawing of the
facility. (b) Zoom on the spherical Couette module, showing the
mountings for the UDV probes and the copper electrodes for
electric potential measurements. $r_i=4.5$\,cm, $r_o=9$\,cm.}
\label{fig:hedgehog_anlage}       
\end{figure}

The terminus spherical Couette flow refers to the fluid 
motion between two  differentially rotating spherical shells.
If the liquid is electrically conducting and exposed to an 
external magnetic field, the set-up is sometimes 
called magnetized spherical Couette flow (MSCF). For resting 
outer cylinder, 
the system is completely defined by three dimensionless 
parameters \cite{Ruediger_2013}:
the Reynolds number ${\rm Re}=\Omega_i r_i^2/\nu$ as a 
measure of the rotation (with angular velocity $\Omega_i$ 
of the inner sphere with radius $r_i$), 
the Hartmann number ${\rm{Ha}}=B_0 r_i \sqrt{\sigma/\rho\nu}$
with $B_0$ denoting the strength of the 
applied axial magnetic field, 
and the geometric  aspect ratio $\eta=r_i/r_o$ with the 
outer sphere radius $r_o$.

\begin{figure}[h]
\includegraphics[width=0.95\textwidth]{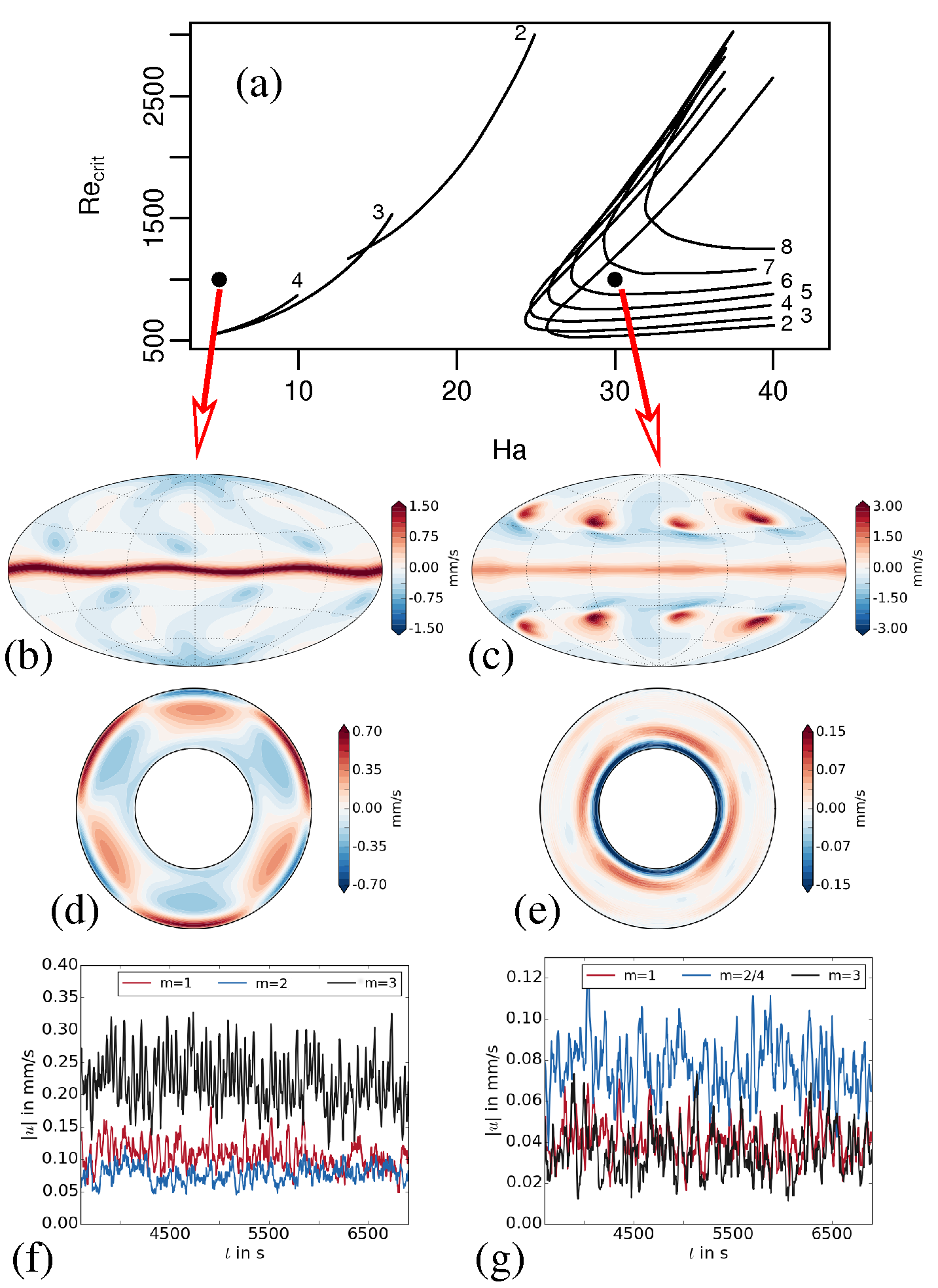}
\caption{Numerical simulations and experimental results
of the HEDGEHOG experiment. (a) Boundaries of instabilities
in dependence on $\rm Ha$ and $\rm Re$, 
reproduced from \cite{Travnikov_2011}. 
Lower left: Jet instability at $\rm Rm=1000$ and $\rm Ha=5$. (b) 
Meridional view of the simulated radial velocity component at 
$r=0.85 r_o$, showing the anti-symmetric character with respect to the
equator. (d) Polar view  of the meridional velocity component
at $\theta=\pi/2$, showing an $m=3$ azimuthal dependence. 
(f) Azimuthal Fourier components of the UDV measured velocity
 taken at a depth of 25\,mm, showing the expected dominance
of the $m=3$ mode. 
Lower right: Return flow instability at $\rm Rm=1000$ and $\rm Ha=30$.
(c) Meridional view of the simulated radial velocity component at 
$r=0.6 r_o$, showing the symmetric character with respect to the
equator. (e) Polar view  of the meridional velocity component
at $\theta=\pi/2$, showing an $m=4$ azimuthal dependence. 
(g) Azimuthal Fourier components of the UDV measured velocity
 taken at a depth of 36\,mm, showing the expected dominance
of the $m=4$ mode.
}
\label{fig:travnikov}       
\end{figure}

Besides its paradigmatic character, the MSCF gained 
new attention in the astrophysical community as 
Sisan et al. \cite{Sisan_2004} had claimed the observation of the 
MRI on the background of a fully 
turbulent flow. 
Later, however, Hollerbach \cite{Hollerbach_2009} 
and Gissinger et al. \cite{Gissinger_2011} interpreted
the observed fluctuations just as turbulent analogues of 
various well-known MSCF instabilities.

In order to clarify this point, the new apparatus 
HEDGEHOG ({\it H}ydromagnetic {\it E}xperiment with {\it D}ifferentially 
{\it G}yrating
sph{\it E}res {\it HO}lding {\it G}aInSn) has been 
installed at HZDR (see Fig.~\ref{fig:hedgehog_anlage}).
In contrast to \cite{Sisan_2004}, HEDGEHOG works 
in a quasi-laminar regime for which numerical 
reference data is available for the two particular
aspect ratios $\eta=0.35$ \cite{Hollerbach_2009} and
 $\eta=0.5$ \cite{Travnikov_2011}.

\subsection{Numerical simulations}
\label{subsec:3-1}

The experiment has been numerically simulated by two different 
codes. One is the code described in  \cite{Hollerbach_2000} which 
solves for a flow, driven by the rotating inner sphere, according 
to the incompressible Navier-Stokes Equation
including a Lorentz force term. The steady states of the full 
three-dimensional calculation have the
practical use of guiding diagnostic design and expectations for the
low $\rm Re$ case. They also demonstrate a saturation 
mechanism for the instabilities \cite{Kaplan_2014}.
Complementary to that code, we have also used the MagIC code 
\cite{Wicht_2014}, which has a long and very successful record
of simulating dynamos in spherical geometry.

Figure~\ref{fig:travnikov} shows the stability boundaries
and illustrates two typical instabilities, 
for a MSCF with aspect ratio $\eta=0.5$. 
First, Fig.~\ref{fig:travnikov}a
shows the  boundaries for the different types 
of instability  in dependence on $\rm Ha$. Actually, the 
lines have been replotted from \cite{Travnikov_2011}, although
large parts of the diagram were reproduced by 
Hollerbach's code. The two full circles represent the two 
distinct types of instability which are separated by a region of 
stability. Their spatial character, as simulated by the 
MagIC code, is illustrated in Fig.~\ref{fig:travnikov}b-e.

At low ${\rm Ha}=5$, the instability arises in the equatorial 
jet, see Fig.~\ref{fig:travnikov}b,d. 
It is anti-symmetric with respect to the
equator, and characterized by an $m=3$ azimuthal dependence.
At higher ${\rm Ha}=30$, the instability arises along 
the Shercliff layer, see Fig.~\ref{fig:travnikov}c,e, 
it is symmetric with respect to the
equator, and has an $m=4$ azimuthal dependence.

\subsection{The HEDGEHOG experiment}
\label{subsec:3-2}

HEDGEHOG (see Fig.~\ref{fig:hedgehog_anlage}) 
consists of one of two optional 
inner spheres ($r_i = 3$ cm or 4.5 cm)
held in the center of an outer sphere ($r_o=9$ cm). The 
outer sphere is a Polymethyl Methacetate
(PMMA) acrylic with 30 
cylindrical holders for ultrasonic 
Doppler velocimetry (UDV), and 168
copper electrodes for electric potential measurement. 

The space between the spheres is filled with
GaInSn. Because of the
high density of this medium, each optional 
inner sphere holds a lead weight to counter the 
buoyancy force. The axial magnetic 
field is provided by a pair of copper 
electromagnets with central radii of 30 cm, 
with a vertical gap of 31 cm between
them (a near Helmholtz configuration). The  
spheres can be  driven with a minimum
rotation frequency $f_i=0.02$ Hz by  90 W 
electromotors.

The data presented here is based on the velocity acquired from 
six equally spaced UDV sensors  on the northern hemisphere (NH), 
and from one UDV sensor on the southern hemisphere (SH).
This sensor configuration allows to identify (roughly) the 
azimuthal dependence of the modes and their symmetry properties 
with respect to the equator.
 
For low $\rm Ha$ the UDV sensors show an outward directed 
velocity around the depth of the equatorial plane.
This is the radial jet whose instability is known from the 
purely hydrodynamic case \cite{Hollerbach_2006}.
Increasing $\rm Ha$ further, the jet is suppressed and a 
growing area with zero velocity values comes up.
At still higher $\rm Ha$ new velocity fluctuations 
occur at greater depth of the UDV beam.

Using 6 sensor data in azimuthal direction,
according to the Nyquist-Shannon criterion a maximum 
wave number $m=3$ can be resolved by Fourier transform. 
Further to this, the frequency  of the equatorially 
symmetric and anti-symmetric velocity parts 
can be inferred from spectrograms.
Using the data from the facing NH and SH
UDV pair, the equatorial symmetric part is computed 
as $(u_{\rm{NH}}+u_{\rm{SH}})/2$, and the anti-symmetric 
part as $(u_{\rm{NH}}-u_{\rm{SH}})/2$.
Figure~\ref{fig:travnikov}f shows the Fourier transform
of the anti-symmetric part taken at a depth of 25\,mm
For this low $\rm Ha=5$ we clearly recognize a 
dominating azimuthal $m=3$ mode, also with the 
numerically predicted frequency (not shown  here).

At $\rm Ha=30$ the instability is distinctly shifted 
towards the inner sphere and acquires an 
equatorially symmetric 
character, which indicates the onset of the
return flow instability.
Figure~\ref{fig:travnikov}g shows the Fourier transform
of the symmetric part taken at a depth of 36\,mm.
We clearly recognize a 
dominating azimuthal $m=4$ mode, also with the 
numerically predicted frequency.

The wave number and frequency observations at 
the chosen $\rm Re=1000$ and increasing values of 
$\rm Ha$ are in good 
agreement with numerical predictions.
In future, electric potential measurements will help to 
resolve the remaining ambiguities with respect 
to the azimuthal wave number of the observed modes. 

\section{Conclusions and Outlook}
\label{sec:4}

Despite the significant theoretical and experimental achievements
on the various inductionless forms of shear or current-driven 
instabilities, the unambiguous laboratory proof of SMRI 
is still elusive. In the following we will sketch our plans 
for a large-scale liquid sodium experiment which is supposed to allow for
studying the transition between the inductionless versions such as
HMRI and AMRI to SMRI. 

Further to this, we will also delineate the prospects for laboratory
studies of instabilities of rotating flows with positive shear.

\subsection{The large MRI/TI experiment}
\label{subsec:4-1}

The DREsden Sodium facility for DYNamo and 
thermohydraulic studies (DRESDYN) is intended as a platform 
both for large-scale experiments related to geo- and astrophysics
as well as for experiments related to thermohydraulic and safety 
aspects of liquid metal batteries and liquid metal fast reactors. 
The most ambitious project in the framework of
DRESDYN is a homogeneous hydromagnetic dynamo driven solely by precession
\cite{Stefani_2012,Stefani_2015}.

\begin{figure}[h]
\includegraphics[width=0.95\textwidth]{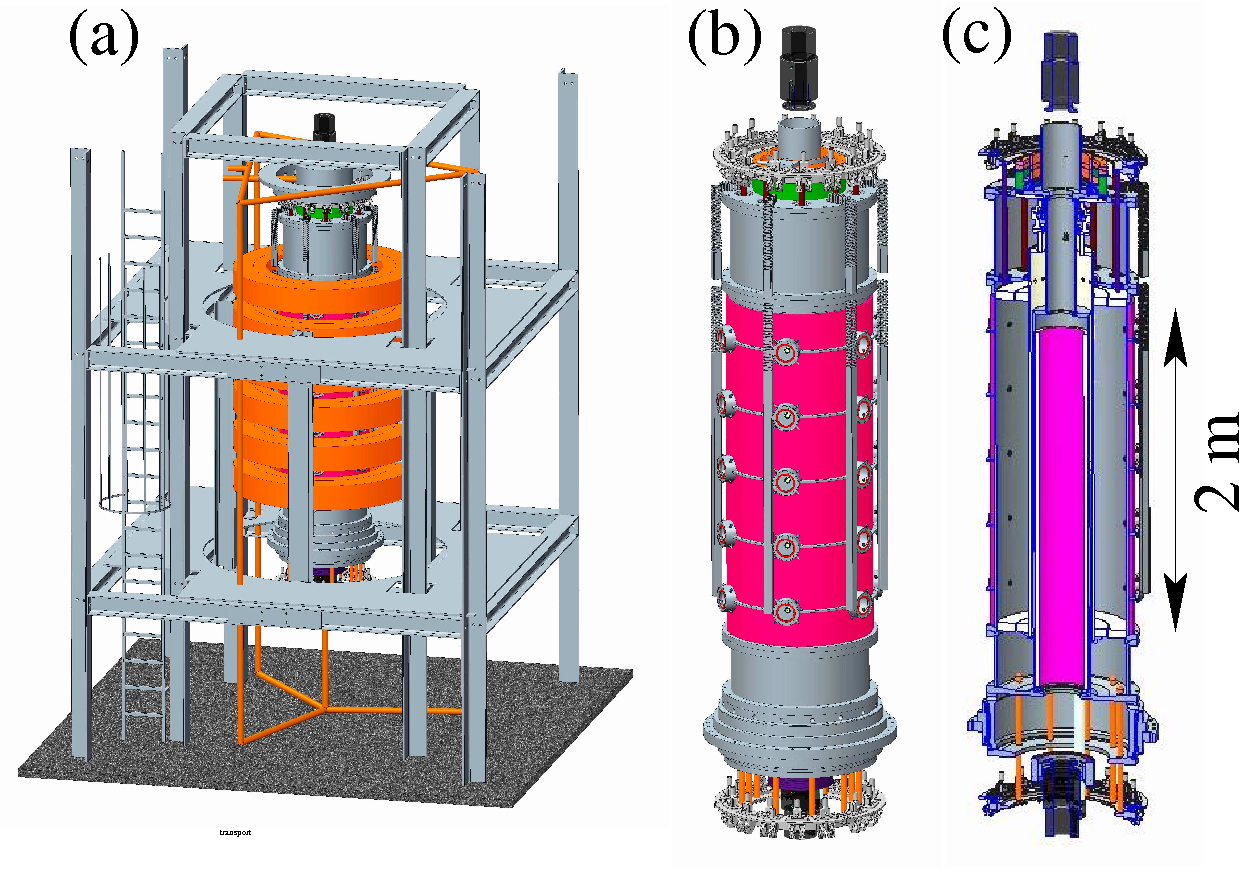}
\caption{Drawings of the combined MRI/TI experiment as planned
in the framework of the DRESDYN project. (a) Total view. (b) Central
module from outside. (c) Central module from inside.}
\label{fig:mriti}   
\end{figure}

A second large-scale experiment relevant to geo- and astrophysics will 
investigate combinations of
different versions of the MRI with the current-driven
TI (see Fig. \ref{fig:mriti}). Basically, the set-up is designed as 
a Taylor-Couette experiment with 2\,m
height, an inner radius $r_i =20$\,cm and an outer radius 
$r_o =40$ cm. 
Rotating the inner cylinder at up to 20\,Hz, we plan to reach 
${\rm Rm} \sim 40$, while the maximum 
axial magnetic field $B_z=120$\,mT will 
correspond to a Lundquist number 
$S:={\rm Pm}^{1/2}{\rm Ha} \sim 8$.
Both values are about twice their respective 
critical values for SMRI as
they were derived in \cite{Ruediger_2003}.

Below those critical values, 
we plan to investigate how the 
helical version of MRI approaches the limit 
of SMRI \cite{Hollerbach_2005}. 
To this end, we will use a
strong central current, as it was already done 
in the PROMISE experiment \cite{Seilmayer_2014,Stefani_2009}.
This insulated central current can be supplemented 
by another axial current guided
through the (rotating) liquid sodium, which will then 
allow to investigate arbitrary
combinations of MRI and TI. As discussed in Sec.~\ref{sec:2}, 
theoretical studies 
\cite{Kirillov_2013,Kirillov_2014}
have shown that even a slight addition of current through 
the liquid would extend the range of application of the 
helical and azimuthal MRI to Keplerian flow profiles.

\subsection{Positive shear instabilities}
\label{subsec:4-2}
The last remarks refer to flows with 
positive ${\rm Ro}$, i.\,e. flows whose
angular velocity (not only the angular momentum) is
increasing outward. From the purely hydrodynamic point of view, such
flows are linearly stable (while non-linear 
instabilities were actually observed in experiments \cite{Tsukahara_2010}). 
Flows with positive ${\rm Ro}$ are indeed relevant
for the equator-near strip (approximately between $\pm 30^{\circ}$) 
of the solar tachocline \cite{Parfrey_2007},
which is, interestingly, also the region of sunspot activity 
\cite{Charbonneau_2010}.

As discussed in Sec.~\ref{sec:2}, a first hint on 
a magnetic destabilization of flows with ${\rm Ro}>{\rm Ro}_{\rm ULL}$
was given in the paper 
\cite{Liu_2006} for the case of a helical magnetic field.
For a purely azimuthal field, 
a similar effect was found both in the framework of
WKB approximation \cite{Stefani_2015b}
as well as by a 1D modal stability analysis \cite{Ruediger_2016}.

What are the prospects for an experiment that could show 
a magnetic destabilization of positive shear flows?
The WKB analysis in \cite{Stefani_2015} has revealed 
the need for a rather narrow gap flow which necessarily 
leads to comparable high axial currents. For a prospective 
Taylor-Couette experiment with Na with an 
outer diameter of $r_o=0.25$\,m, 
and choosing ${\eta}=0.85$, those values would be
${\rm {Re}}=1898$,  ${\rm {Ha}}=892$
which amounts to physical values of $f_o=0.022$\,Hz, 
and $B_{\phi}(r_i)=77$\,mT, requiring
a central current of 82\,kA !

The 1D simulations of \cite{Ruediger_2016} 
predicted more optimistic values in the order of
30 kA, if the radial boundaries were considered as
ideally conducting. It remains to be shown by careful 
numerical studies what are the real critical currents 
for an optimized laboratory implementation when realistic
conductivity ratios between the liquid (sodium) and the wall
material (copper) are taken into account.

\begin{acknowledgement}
This work was supported by Deutsche 
Forschungsgemeinschaft in frame of the focus programme 1488 (PlanetMag).
Intense collaboration with Rainer Hollerbach on the theory and
numerics of the different instabilities is gratefully acknowledged.
We thank Thomas Gundrum for his efforts in setting up 
and running the experiments, and Elliot Kaplan for his
numerical and experimental work on the HEDGEHOG experiment. 
We are grateful to Johannes Wicht for the introduction into 
the MagIC code. F.S. likes to thank Oleg Kirillov
for his effort to establish a comprehensive WKB theory of the 
magnetically triggered instabilities, and George Mamatsashvili
for his work on non-modal aspects of MRI.
\end{acknowledgement}

\begin{thebibliography}{99.}%
%
%

\bibitem{Adams_2015}
Adams, M.M., Stone, D.R., Zimmerman, D.S., Lathrop, D.P.: 
Liquid sodium models of the Earth's core. 
Prog. Earth Planet. Sci. \textbf{29}, 1-18 (2015).

\bibitem{Balbus_2003}
Balbus, S.A.: Enhanced angular momentum transport 
in accretion disks. Ann. Rev. Astron. Astrophys. \textbf{41}, 
555-597 (2003)

\bibitem{Benzi_2010}
Benzi, R., Pinton, J.-F.: 
Magnetic Reversals in a Simple Model of Magnetohydrodynamics. 
Phys. Rev. Lett. \textbf{105}, 
024501 (2010)

\bibitem{Bergerson_2006}
Bergerson, W.F., Hannum, D.A., Hegna, C.C., Kendrick, R.D.,
Sarff, J.S., Forest, C.B.: Onset and saturation of 
the kink instability in a current-carrying line-tied plasma. 
Phys. Rev. Lett. \textbf{96}, 
015004 (2006).

\bibitem{Berhanu_2010}
Berhanu, M. et al.: Dynamo regimes and transitions in the VKS experiment. 
Eur. Phys. J. B \textbf{77}, 459--468 (2010)

\bibitem{Chandra_1956}
Chandrasekhar, S.:
On the stabiluty of the simplest soluation of the
equations of hydromagnetics.
P. Natl. Acad. Sci. USA \textbf{42}, 273-276 (1956)


\bibitem{Charbonneau_2010}
Charbonneau, P.: Dynamo models of the solar cycle.
Liv. Rev. Sol. Phys. \textbf{7}, 3 (2010)

\bibitem{Cooper_2014}
Cooper, C.M. et al.: The Madison plasma dynamo experiment: 
A facility for studying laboratory plasma astrophysics. 
Phys. Plasmas \textbf{21}, 013505 (2014)

\bibitem{Dormy_2016}
Dormy, E.: Strong-field spherical dynamos.
J. Fluid Mech. \textbf{789}, 500-513  (2016) 


\bibitem{Gailitis_2000}
Gailitis, A., Lielausis, O., Dement'ev, S., 
Platacis, E., Cifersons, A., Gerbeth, G., 
Gundrum, T., Stefani, F., Christen, M., 
H\"anel, H., Will, G.: Detection of a 
flow induced magnetic field eigenmode in the Riga dynamo facility.
Phys. Rev. Lett. \textbf{84}, 4365--4369 (2000)

\bibitem{Gailitis_2002}
Gailitis, A., Lielausis, O., PLatacis, E., Gerbeth, G.,
Stefani, F.: Laboratory experiments on hydromagnetic dynamos,
Rev. Mod. Phys.  \textbf{74}, 973--990  (2002)

\bibitem{Gellert_2007}
Gellert, M., R\"udiger, G., Fournier, A.: 
Energy distribution in nonaxisymmetric magnetic Taylor-Couette 
flow. 
Astron. Nachr. \textbf{328}, 1162-1165 (2007)


\bibitem{Gellert_2008}
Gellert, M., R\"udiger, G., Elstner, D.: Helicity 
generation and alpha-effect by Tayler instability 
with z-dependent differential rotation. 
Astron. Astrophys. \textbf{479}, L33--L36 (2008)




\bibitem{Giesecke_2010}
Giesecke, A., Stefani, F., Gerbeth, G.: 
Role of soft-iron impellers on the mode selection in 
the von-Karman-sodium dynamo experiment. 
Phys. Rev. Lett. \textbf{104}, 044503 (2010)


\bibitem{Giesecke_2012}
Giesecke, A., Nore, C., Stefani, F., 
Gerbeth, G., L\'eorat, J., Herreman, W., 
F., Guermond, J.-L.: Influence of high-permeability 
discs in an axisymmetric model of the Cadarache 
dynamo experiment. New. J. Phys. \textbf{14}, 053005 (2012)


\bibitem{Gissinger_2011}
Gissinger, C., Ji, H., Goodman, J.: Instabilities in 
magnetized spherical Couette flow. Phys. Rev. E \textbf{84}, 
026308   (2011)

\bibitem{Hollerbach_2000}
Hollerbach, R.:
A spectral solution of the magneto-convection 
equations in spherical geometry.
Int. J. Num. Meth. Fluids \textbf{32},
773--797 (2000)


\bibitem{Hollerbach_2005}
Hollerbach, R., R\"udiger, G.: New type of 
magnetorotational instability in cylindrical 
Taylor-Couette flow. Phys. Rev. Lett. {\bf 95}, 124501 (2005)

\bibitem{Hollerbach_2006}
Hollerbach, R.: Non-axisymmetric instabilities in 
basic state spherical Couette flow.
Fluid. Dyn. Res.  \textbf{38},
257--273  (2006)

\bibitem{Hollerbach_2009}
Hollerbach, R.: Non-axisymmetric instabilities 
in magnetic spherical Couette flow. 
Proc. R. Soc. A \textbf{465}, 2003--2013 (2009).

\bibitem{Hollerbach_2010}
Hollerbach, R., Teeluck, V., R\"udiger, G.: 
Nonaxisymmetric magnetorotational instabilities in cylindrical 
Taylor-Couette flow. Phys. Rev. Lett. \textbf{104}, 044502 (2010).


\bibitem{Jones_2011}
Jones, C.A.: Planetary magnetic fields and dynamos. 
Ann. Rev. Fluid Mech. \textbf{43}, 583--614 (2011)

\bibitem{Kaplan_2014}
Kaplan, E.: Saturation of nonaxisymmetric 
instabilities of magnetized spherical Couette flow.
Phys. Rev. E. \textbf{89}, 063016  (2014) 

\bibitem{Kim_2012}
Kim, H. et al.: Liquid metal batteries: past, present, and future.
Chem. Rev. \textbf{113}, 2075--2099 (2013)


\bibitem{Kirillov_2010}
Kirillov, O.N.; Stefani, F.:
On the relation of standard and helical magnetorotational
instability. Astrophys. J. \textbf{712}, 
52--68  (2010) 

\bibitem{Kirillov_2011}
Kirillov, ON.; Stefani, F.:
Paradoxes of magnetorotational instability and their 
geometrical resolution.
Phys. Rev. E \textbf{84},  036304  (2011)

\bibitem{Kirillov_2012}
Kirillov, O.N., Stefani, F., 
Fukumoto, Y.:
A unifying picture of helical and azimuthal 
magnetorotational instability, and the universal significance
of the Liu limit. Astrophys. J. \textbf{756}, 
83 (2012) 

\bibitem{Kirillov_2013}
Kirillov, O.N., Stefani, F.:
Extending the range of the inductionless 
magnetorotational instability.
Phys. Rev. Lett. \textbf{111}, 
061103  (2013)

\bibitem{Kirillov_2014}
Kirillov, O.N., Stefani, F., 
Fukumoto, Y.:
Local instabilities in magnetized rotational flows: 
a short-wavelength approach.
J. Fluid Mech. \textbf{760}, 591--633 (2014) 


\bibitem{Lathrop_Forest_2011}
Lathrop, D.P., Forest, C.B.: 
Magnetic dynamos in the lab. 
Phys. Today \textbf{64}, 40--45 (2011)

\bibitem{Lebreton_1987}
Lebreton, Y., Maeder, A.: 
Stellar evolution with turbulent diffusion mixing. VI - The solar model, 
surface Li-7, and He-3 abundances, solar neutrinos and oscillations.
Astron. Astrophys. \textbf{175}, 99 (1987)


\bibitem{Liu_2006}
Liu, W., Goodman, J., Herron, I., Ji, H.: 
Helical magnetorotational instability in magnetized 
Taylor-Couette flow. Phys. Rev. E \textbf{74}, 
056302 (2006)


\bibitem{Liu_2007}
Liu, W., Goodman, J., Ji, H.: Traveling waves in a 
magnetized Taylor-Couette flow. Phys. Rev. E \textbf{76}, 
016310 (2007)


\bibitem{MAGIC}
https://github.com/magic-sph/magic.



\bibitem{Mama_2016}
Mamatsashvili, G., Stefani, F.:
Linking dissipation-induced instabilities 
with nonmodal growth: the case of helical magnetorotational 
instability. arXiv:1604.07205

\bibitem{Moll_2008}
Moll. R., Spruit, H.C., Obergulinger, M.:
Kink instabilities in jets from 
rotating magnetic fields.
Astron. Astrophys. \textbf{492}, 
621--630  (2008)

\bibitem{Montgomery_1993}
Montgomery, D.: Hartmann, Lundquist, and Reynolds: the role of 
dimensionless numbers in nonlinear magnetofluid behavior.
Phys. Rev. E \textbf{87}, 012108 (2013)

\bibitem{Mori_2013}
Mori, N. Schmitt, D., Wicht, J. 
Ferriz-Mas, A., Mouri, H., 
Nakamichi, A. Morikawa, M: Domino model for geomagnetic field reversals.
Plasma Phys. Control. Fusion \textbf{35}, 105--113 (1993)

\bibitem{Nore_2016}
Nore, C., Quiroz, D.C., Cappanera, L., Guermond, J.L.:
Direct numerical simulation of the axial dipolar 
dynamo in the Von K\'arm\'an Sodium experiment.
Europhys. Lett. \textbf{114}, 65002 (2016)

\bibitem{Nornberg_2010}
Nornberg, M.D., Ji, H., Schartman, E., Roach, A., 
Goodman, J.: Observation of magnetocoriolis waves in a 
liquid metal Taylor-Couette experiment. 
Phys. Rev. Lett. \textbf{104}, 074501 (2010)

\bibitem{Paredes_2016}
Paredes, A., Gellert, M., R\"diger, G.:
Mixing of a passive scalar by the instability 
of a differentially rotating axial pinch.
Adtron. Astrophys.  \textbf{588}, A147 (2016)



\bibitem{Parfrey_2007}
K.P. Parfrey, K. Menou:
The origin of solar activity in the tachocline:
Astrophys. J. Lett. \textbf{667}, L207
(2007)

\bibitem{Petitdemange_2008}
Petitdemange, L., Dormy, E., Balbus, S.A.: 
Magnetostrophic MRI in the Earth's outer core. 
Geophys. Res. Lett. \textbf{35}, L15305 (2008)

\bibitem{Petitdemange_2010}
Petitdemange, L.: Two-dimensional 
non-linear simulations of the magnetostrophic 
magnetorotational instability. Geophys. Astrophys. 
Fluid Dyn. \textbf{104}, 287--299 (2010)

\bibitem{Priede_2009}
Priede, J., Gerbeth, G.: Absolute versus convective 
helical magnetorotational instability in a Taylor-Couette 
flow. Phys. Rev. E \textbf{79}, 046310 (2009)


\bibitem{Petrelis_2009}
Petrelis, F., Fauve, S., Dormy, E., Valet, J.-P.: 
Simple mechanism for reversals of Earth's 
magnetic field. Phys. Rev. Lett. \textbf{102}, 144503 (2009)

\bibitem{Reuter_2009}
Reuter, K., Jenko, F., Tilgner, A., 
Forest, C.B.: 
Wave-driven dynamo action in spherical 
magnetohydrodynamic systems. Phys. Rev. E \textbf{80}, 
056304 (2009)

\bibitem{Roach_2012}
Roach, A.H., Spence, E.J., Gissinger, C., Edlund, E.M., 
Sloboda, P., Goodman, J., Ji, H.: Observation of a 
free-Shercliff-layer instability in cylindrical geometry. 
Phys. Rev. Lett. \textbf{108}, 154502 (2012)

\bibitem{Ruediger_2003}
R\"udiger, G., Shalybkov, D.: 
Linear magnetohydrodynamic Taylor-Couette instability 
for liquid sodium. Phys. Rev. E \textbf{67}, 
046312  (2003) 

\bibitem{Ruediger_2005}
R\"udiger G., Hollerbach R., Schultz M., Shalybkov D.:
The stability of MHD Taylor-Couette flow with 
current-free spiral magnetic fields between conducting cylinders.
Astron. Nachr. \textbf{326}, 409-413 (2005)

\bibitem{Ruediger_2007}
R\"udiger, G., Hollerbach, R., Schultz, M., Elstner, D.: 
Destabilization of hydrodynamically stable rotation laws by azimuthal
magnetic fields. Mon. Not. R. Astron. Soc. \textbf{377}, 
1481--1487  (2007)

\bibitem{Ruediger_2009}
R\"udiger, G., Gellert, M., Schultz, M.: 
Eddy viscosity and turbulent Schmidt number by kink-type 
instabilities of toroidal magnetic fields. 
Mon. Not. R. Astron. Soc. \textbf{399}, 
996--1004 (2009)

\bibitem{Ruediger_2011}
R\"udiger, G., Schultz, M., Gellert, M.:
The Tayler instability of toroidal magnetic fields in a columnar gallium experiment. Astron. Nachr. \textbf{332}, 17-23 (2011) 



\bibitem{Ruediger_2013}
R\"udiger, G., Hollerbach, R., Kitchatinov, L.L.:
Magnetic processes in astrophysics: theory, simulations, experiments. 
WILEY-VCH, Weinheim (2013)

\bibitem{Ruediger_2014}
R\"udiger, G., Gellert, M., Schultz, M., Hollerbach, R., Stefani, F.:
Astrophysical and experimental implications from the
magnetorotational instability of toroidal fields.
Mon. Not. R. Astron. Soc. \textbf{438}, 271-277 (2014)


\bibitem{Ruediger_2015}
R\"udiger, G., Schultz, M., Stefani, F., Mond. M.:
Diffusive magnetohydrodynamic instabilities beyond 
the Chandrasekhar theorem.
Astrphys. J. \textbf{811}, 84 (2015)


\bibitem{Ruediger_2016}
R\"udiger, G., Schultz, M., Gellert, M., Stefani, F.:
Subcritical excitation of the 
current-driven Tayler instability by super-rotation.
Phys. Fluids \textbf{28}, 014105 (2016)

\bibitem{Schatzmann_1977}
Schatzman, E.:
Turbulent transport and lithium destruction in main sequence stars.
Astron. Astrophys. \textbf{56}, 211 (1977)

\bibitem{Schmitt_2013}
Schmitt, D., Cardin, P., La Rizza, P., Nataf, H.C.:
Magneto-Coriolis waves in a spherical Couette flow experiment. 
Eur. J. Phys. B - Fluids \textbf{37}, 
10--22 (2013)

\bibitem{Seilmayer_2012}
Seilmayer, M., Stefani, F., Gundrum, T., 
Weier, T., Gerbeth, G.:
Experimental evidence for a transient 
Tayler instability in a cylindrical liquid-metal column.
Phys. Rev. Lett. \textbf{108}, 244501 (2012)



\bibitem{Seilmayer_2014}
Seilmayer, M., Galindo, V., Gerbeth, G., Gundrum, T., Stefani, F., 
Gellert, M., R\"udiger, G., Schultz, M.:
Experimental evidence for nonaxisymmetric 
 magnetorotational instability in a rotating liquid metal 
 exposed to an azimuthal magnetic field.
 Phys. Rev. Lett. \textbf{113}, 024505 (2014) 


\bibitem{Seilmayer_2016}
Seilmayer, M., Gundrum, T., Stefani, F.:
Noise reduction of ultrasonic Doppler velocimetry in 
liquid metal experiments with high magnetic fields.
Flow Meas. Instrum. \textbf{48}, 74--80 (2016) 

\bibitem{Sisan_2004}
Sisan, D.R., Mujica, N., Tillotson, W.A., Huang, Y.M,
Dorland, W., Hassam, A.B., Lathrop, D.P.: 
Experimental observation and characterization of the magnetorotational 
instability. Phys. Rev. Lett. \textbf{93}, 114502 (2004)

\bibitem{Sorriso_2007}
Sorriso-Valvo, L., Stefani, F., Carbone, V. 
Nigro, G., 
Lepreti, F.,  Vecchio, A. Veltri, P:
A statistical analysis of polarity reversals of the geomagnetic field.
Phys. Earth Planet. Inter.  \textbf{164}, 197-207 (2007)

\bibitem{Spada_2016}
Spada, F., Gellert, M., Arlt, R., Deheuvels, S.: 
Angular momentum transport efficiency in post-main sequence low-mass stars.
Astron. Astrophys. \textbf{589}, A23  (2016) 

\bibitem{Spies_1988}
Spies,  G.O.:  
Visco-resistive stabilization of kinks with short 
wavelengths along an elliptic magnetic stagnation line.
Plasma Phys. Control. Fusion \textbf{30}, 1025--1037 (1988)

\bibitem{Spruit_2002}
Spruit, H.C.: Dynamo action by differential rotation in a 
stably stratified stellar interior. 
Astron. Astrophys. \textbf{381}, 923--932 (2002).


\bibitem{Sreenivasan_2011}
Sreenivasan, B., Jones, C.A.: 
Helicity generation and subcritical behaviour 
in rapidly rotating dynamos. J. Fluid Mech. \textbf{688}, 
5--30 (2011)


\bibitem{Starace_2015}
Starace, M., Weber, N., Seilmayer, M., 
Kasprzyk, C., Weier, T., Stefani, F., Eckert, S.:
Ultrasound Doppler flow measurement in a liquid metal 
colums under the influence of a strong axial 
electric current.
Magnetohydrodynamics \textbf{51}, 249--256 (2015)

\bibitem{Stefani_2006a}
 Stefani, F., Gerbeth, G., G\"unther, U., Xu, M:
 Why dynamos are prone to reversals.
 Earth Planet. Sci. Lett. \textbf{243}, 828--840 (2006)
 

\bibitem{Stefani_2006}
Stefani, F., Gundrum, T., Gerbeth, G., R\"udiger, G., 
Schultz, M., Szklarski, J., Hollerbach, R.:
Experimental evidence for magnetorotational instability 
in a Taylor-Couette flow under the influence of a helical
magnetic field. Phys. Rev. Lett. \textbf{97}, 184502 (2006)

\bibitem{Stefani_2007}
Stefani, F., Gundrum, T., Gerbeth, G., R\"udiger, G., 
Szklarski, J., Hollerbach, R.:
Experiments on the magnetorotational instability in 
helical magnetic fields. New J. Phys. \textbf{9}, 295 (2007)



\bibitem{Stefani_2008}
Stefani, F., Gailitis, A., Gerbeth, G.: 
Magnetohydrodynamic experiments on cosmic magnetic fields. 
Zeitschr. Angew. Math. Mech. \textbf{88}, 930--954 (2008)

\bibitem{Stefani_2009}
Stefani, F., Giesecke, A., Gerbeth, G.: 
Numerical simulations of liquid metal experiments 
on cosmic magnetic fields. 
Theor. Comp. Fluid Dyn. \textbf{23}, 405--429 (2009).

\bibitem{Stefani_2009a}
Stefani, F., Gerbeth, G., Gundrum, T., Hollerbach, R., 
Priede, J., R\"udiger, G., 
Szklarski, J.:
Helical magnetorotational instability in a 
Taylor-Couette flow with strongly reduced Ekman pumping. 
Phys. Rev. E \textbf{80}, 066303 (2009)


\bibitem{Stefani_2011}
Stefani, F.,  Weier, T., Gundrum, T., Gerbeth, G.:
How to circumvent the size 
limitation of liquid metal batteries due to the Tayler instability.
Energy Conv. Manag.  \textbf{52}, 2982--2986  (2011)


\bibitem{Stefani_2012}
Stefani, F., Eckert, S., 
Gerbeth, G., Giesecke, A., Gundrum, T., 
Steglich, C., Wustmann, B.: DRESDYN - 
A new facility for MHD experiments with liquid sodium. 
Magnetohydrodynamics \textbf{48}, 103--113 (2012)

\bibitem{Stefani_2015}
Stefani, F., Albrecht, T., 
Gerbeth, G., Giesecke, A., Gundrum, T., Herault, J., Nore, C.
Steglich, C.: Towards a precession driven dynamo experiement. 
Magnetohydrodynamics \textbf{51}, 275--284 (2015)

\bibitem{Stefani_2015b}
Stefani, F., Kirillov, O.N.: 
Destabilization of 
rotating flows with positive shear by azimuthal magnetic fields.
Phys. Rev. E \textbf{92}, 051001  (2015)

\bibitem{Stefani_2016a}
Stefani, F., Giesecke, A., Weber, N., Weier, T.: 
Synchronized helicity oscillations: A link between 
planetary tides and the solar cycle?. 
Solar Phys. \textbf{291}, 2197-2212 (2016)

\bibitem{Stefani_2016}
Stefani, F., Galindo, V., Kasprzyk, C., Landgraf, S., 
Seilmayer, M., Starace, M., Weber, N., Weier, T.:
Magnetohydrodynamic effects in liquid metal batteries.
IOP Conf. Ser.: Mater. Sci. Eng. \textbf{143}, 012024  (2016)


\bibitem{Szklarski_2007}
Szklarski, J.:
Reduction of boundary 
effects in the spiral MRI experiment PROMISE.
Astron. Nachr.  \textbf{328}, 499--506  (2007) 

\bibitem{Stieglitz_2001}
Stieglitz, R., M\"uller, U.: 
Experimental demonstration of a homogeneous two-scale dynamo. 
Phys. Fluids \textbf{13}, 561--564 (2001)

\bibitem{Tayler_1973}
Tayler, R.J.: Adiabatic stability of stars containing magnetic 
fields. 1 Toroidal fields. Mon. Not. R. Astron. Soc. 
\textbf{161}, 365--380 (1973)


\bibitem{Tilgner_2008}
Tilgner, A.: 
Dynamo action with wave motion. 
Phys. Rev. Lett. {\textbf 100}, 128501 (2008)


\bibitem{Travnikov_2011}
Travnikov, V., Eckert, K., Odenbach, S.:
Influence of an axial magnetic 
field on the stability of spherical Couette 
flows with different gap widths.
Acta Mech. {\textbf 219}, 255--268  (2011) 


 

\bibitem{Tsukahara_2010}
Tsukahara, T., Tillmark, N., Alfredsson,P.H.:
Flow regimes in a plane Couette flow with system rotation.
J. Fluid Mech. {\textbf 648}, 5--33  (2010) 

\bibitem{Weber_2013}
Weber, N., Galindo, V., Stefani, F., Weier, T.:
Numerical simulation of the Tayler instability in liquid metals.
{\textbf 15}, 043034 (2013) 

\bibitem{Weber_2014}
Weber, N., Galindo, V.,  
Stefani, F., Weier, T.:
Current-driven flow instabilities in 
large-scale liquid metal batteries, and how to tame them.
J. Power Sources {\textbf 265}, 166--173 (2014) 

\bibitem{Weber_2015}
 Weber, N., Galindo, V.,  
Stefani, F., Weier, T.:
 The Tayler instability at low magnetic Prandtl numbers: 
 between chiral symmetry breaking and helicity oscillations.
New J. Phys. {\textbf 17}, 113013 (2015) 


\bibitem{Wicht_Tilgner_2010} 
Wicht J., Tilgner, A.: Theory and modeling of planetary dynamos. 
Space Sci. Rev. \textbf{152}, 501--542 (2010)

\bibitem{Wicht_2014} 
Wicht J.: Flow instabilities in the wide-gap spherical Couette system. 
J. Fluid Mech. \textbf{738}, 184--221 (2014)

\bibitem{Zahn_1990}
Zahn, J. P.: in  Rotation and Mixing in Stellar Interiors, 
ed.  M.-J. Goupil, \& J.-P. Zahn, 
Lecture Note of Physics 336 (Springer Verlag) p. 141 (1990)

\bibitem{Zimmermann_2014} 
Zimmermann, D.S., Triana, S.A., Nataf, H.-C., Lathrop, D.P.: 
A turbulent, high magnetic Reynolds number experimental model 
of Earth's core. 
J. Geophys. Res. - Sol. Earth \textbf{119}, 4538--4557 (2010)
 
\end{thebibliography}
\end{document}